\documentclass[sigconf]{acmart}

\acmYear{2027}
\copyrightyear{2027}

\usepackage{enumitem}
\usepackage[most]{tcolorbox}
\usepackage{subcaption}
\usepackage{multirow}
\usepackage{colortbl}
\usepackage{algorithm}
\usepackage{algpseudocode}


\definecolor{headercolor}{gray}{0.94}
\definecolor{bestcolor}{RGB}{221,235,247}
\definecolor{secondcolor}{RGB}{252,228,214}
\newcommand{\bestscore}[1]{\cellcolor{bestcolor}{\textbf{#1}}}
\newcommand{\secondscore}[1]{\cellcolor{secondcolor}{\underline{#1}}}

\newcommand{\TEXTset}{\texttt{TEXT}}
\newcommand{\VISIONset}{\texttt{VISION}}
\newcommand{\FUSIONset}{\texttt{FUSION}}

\title{VLM2Rec: Resolving Modality Collapse in Vision-Language Models for Sequential Recommendation}

\author{Junyoung Kim}
\email{junyoungkim@postech.ac.kr}
\affiliation{%
  \institution{Pohang University of \\ Science and Technology}
  \city{Pohang}
  \country{Republic of Korea}
}
\author{Woojoo Kim}
\email{kimuj0103@postech.ac.kr}
\affiliation{%
  \institution{Pohang University of \\ Science and Technology}
  \city{Pohang}
  \country{Republic of Korea}
}
\author{Jaehyung Lim}
\email{jaehyunglim@postech.ac.kr}
\affiliation{%
  \institution{Pohang University of \\ Science and Technology}
  \city{Pohang}
  \country{Republic of Korea}
}
\author{Dongha Kim}
\email{dhkim0317@postech.ac.kr}
\affiliation{%
  \institution{Pohang University of \\ Science and Technology}
  \city{Pohang}
  \country{Republic of Korea}
}
\author{Wonbin Kweon}
\email{wonbinkweon@skku.edu}
\authornote{Corresponding authors.}
\affiliation{%
  \institution{Sungkyunkwan University}
  \city{Suwon}
  \country{Republic of Korea}
}
\author{Hwanjo Yu}
\email{hwanjoyu@postech.ac.kr}
\authornotemark[1]
\affiliation{%
  \institution{Pohang University of \\ Science and Technology}
  \city{Pohang}
  \country{Republic of Korea}
}

\begin{document}
\renewcommand{\shortauthors}{Junyoung Kim et al.}
\begin{abstract}
\label{abstract}
Sequential Recommendation (SR) learns sequence-level Collaborative Filtering (CF) from interactions, but ID-based models struggle with sparse data and cold-start items.
In contrast, item titles and images can reveal preference continuity, such as a recurring color or style.
Yet existing approaches miss visual continuity: small encoders take a single item as input, and Large Language Models (LLMs) read only titles.
We therefore shift our focus to Large Vision-Language Models (VLMs), which can read a user's entire image sequence, not just its titles.
However, VLMs inherently suffer from \textbf{modality collapse}: they over-rely on text and, as a result, overlook visual continuity for recommendation.
We find that standard contrastive Supervised Fine-Tuning (SFT), essential for adapting VLMs to SR, further amplifies this collapse: updates follow the text gradient, visual discrimination weakens, and adding image embeddings to the text embeddings consistently lowers accuracy.
We propose \textbf{VLM2Rec} with two objectives: \textit{Weak-modality Penalized Contrastive Learning} strengthens each user's weaker modality, and \textit{Cross-modal Relational Topology Regularization} aligns each user's relative similarity structure over the same items across the two modality spaces.
Its variant, $\text{VLM2Rec}_{\text{plug}}$, reuses the learned item embeddings in SR backbones and thus serves at the latency of conventional SR models.
Experiments show that VLM2Rec mitigates the collapse and consistently improves accuracy and robustness.
\end{abstract}

\maketitle

\section{Introduction}
\label{sec:intro}
Sequential Recommendation (SR), which drives next-item suggestions on Web platforms such as e-commerce and streaming services, models evolving user preferences through sequence-level Collaborative Filtering (CF), learning transition patterns that capture which items follow a given interaction history.
Traditional ID-based models~\cite{sasrec,gru4rec,fmlp} learn these transitions from interactions alone, which makes them vulnerable to data sparsity and cold-start items.
Meanwhile, items on these platforms come with rich content, including titles and images.
This content across consecutive items reveals \textit{preference continuity}, such as a recurring color or style, and thus explains why transitions occur.

Content-based SR exploits these contents to capture preference continuity and to alleviate the data sparsity and cold-start problems of ID-based models.
To do so, multimodal SR models~\cite{mmrec1,modrec_m3srec,modrec_hm4sr} feed item features from small pretrained encoders such as BERT~\cite{bert} and CLIP~\cite{clip} into an SR backbone such as SASRec~\cite{sasrec}, yet these encoders view each item in isolation and offer limited semantic understanding.
Recently, Large Language Models (LLMs) read the titles in a history with strong text understanding, capturing textual continuity and substantially improving SR~\cite{slim,llm2rec}.
However, LLMs read only text and thus cannot exploit visual continuity in item images.
Feeding CLIP image features into an LLM~\cite{illmrec} does not fully resolve this limitation, because the LLM interprets these features within its text space and compresses each image into a single token, discarding fine-grained details.

Unlike these LLM-based approaches, Large Vision--Language Models (VLMs)~\cite{qwen2.5-vl,internvl3} directly process raw images and text, which makes them well suited to multimodal SR for three reasons.
\textbf{First}, large-scale vision--language pretraining provides rich visual semantics and cross-modal priors.
\textbf{Second}, VLMs' strong multi-image reasoning lets them jointly compare the images in a user's sequence and capture visual continuity, such as a recurring color or style.
\textbf{Third}, they accept images together with textual instructions, so a prompt such as ``Describe this item image sequence'' can guide them to understand an image history for recommendation.
For these reasons, we apply VLMs to SR so that they read users' multimodal histories and exploit both textual and visual continuity.

However, applying VLMs directly to recommendation leaves a problem inherent to them: \textbf{modality collapse}.
They are known to over-rely on the dominant modality, usually text, while underusing the weaker one~\cite{modality_bias1,modality_gap_gen1}, even as recent studies use them to rank candidate items directly~\cite{mllmmsr,hanorec} or to derive item representations~\cite{notellm2,vlm_prompt}.
Our experiments (Sec.~\ref{sec:empirical}) show that this collapse in vanilla VLMs hampers recommendation: visual embeddings barely separate the next item from negatives, and adding image embeddings to the text embeddings can even lower accuracy (i.e., a negative transfer).
Thus, the collapse undermines the very multimodality that motivates VLMs, leaving visual continuity overlooked for recommendation; resolving it is crucial.

\begin{figure*}[t]
    \centering
    \includegraphics[width=\textwidth]{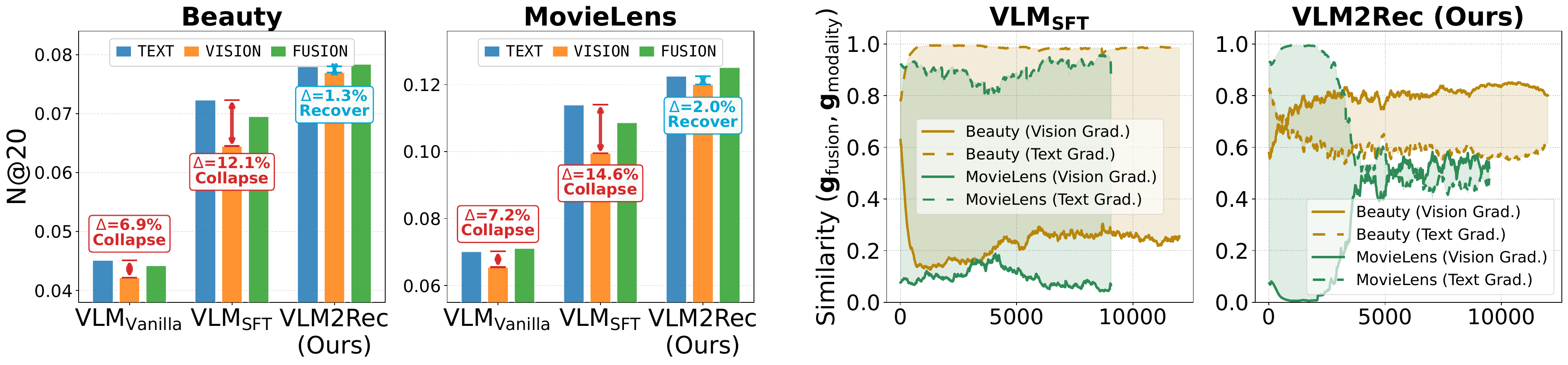}
    \Description{Bar charts compare recommendation performance with text-only, vision-only, and fused inputs on Beauty and MovieLens. Line charts compare the alignment of text and vision gradients with the fused gradient during SFT and VLM2Rec training.}
    \caption{Analysis of modality collapse in the VLM2Rec setting (N@20): dropout test (Left) and gradient dynamics (Right). $\text{VLM}_{\text{SFT}}$ widens the \TEXTset{}--\VISIONset{} gap and favors text gradients; VLM2Rec (ours) narrows it and rebalances updates.}
    \label{fig:drop_grad}
\end{figure*}

Yet we find that standard contrastive Supervised Fine-Tuning (SFT)~\cite{infonce1}, an essential step for adapting VLMs to SR, further amplifies this collapse (Sec.~\ref{sec:empirical}).
SFT updates increasingly follow the text gradient, widening the gradient gap between text and vision, which suggests that the VLM takes the easier text modality as a \textit{shortcut}~\cite{shortcut2} to reduce the loss.
As a result, the visual modality loses even more of its ability to separate each user's next item from negatives, and text-only inputs outperform vision-only inputs by a widening margin.
The negative transfer thus becomes consistent: adding image embeddings to the text embeddings lowers accuracy on both datasets (Fig.~\ref{fig:drop_grad} left).
To fully exploit the cross-modal capabilities of VLMs for recommendation, we need an objective that, unlike standard SFT, corrects this collapse.

To this end, we propose \textbf{VLM2Rec} (Sec.~\ref{sec:methods}), which corrects the collapse with two objectives.
\textit{Weak-modality Penalized Contrastive Learning} ($\mathcal{L}_{\text{WPCL}}$) identifies the weaker modality of each user, since users differ in how much they rely on textual or visual cues, and strengthens it with a penalty weight that grows with its gap from the stronger modality.
\textit{Cross-modal Relational Topology Regularization} ($\mathcal{L}_{\text{CRTR}}$) softly aligns the neighborhood of each user's history in the two modality spaces without forcing identical text and vision embeddings, so that fusing them improves recommendation.
Moreover, we introduce $\text{VLM2Rec}_{\text{plug}}$, which transfers the trained item embeddings to SR backbones and thus serves without VLM calls at the latency of conventional SR models (Sec.~\ref{sec:deployment}).
Experiments (Sec.~\ref{sec:exp}) show that VLM2Rec mitigates the collapse across benchmarks and consistently improves accuracy and robustness.
The main contributions of this paper are as follows:
\begin{itemize}[leftmargin=*]
    \item We empirically find that standard contrastive SFT, though essential for adapting VLMs to SR, amplifies their modality collapse by under-optimizing the weaker modality.
    \item We propose VLM2Rec, which strengthens each user's weaker modality and aligns each user's relative similarity structure in the two modality spaces; its variant $\text{VLM2Rec}_{\text{plug}}$ enhances SR backbones with the item embeddings while matching the serving latency of conventional SR models.
    \item We demonstrate the effectiveness of VLM2Rec through extensive experiments across datasets, SR backbones, and VLM families.
\end{itemize}

\section{Related Work}
\label{sec:related}
\noindent\textbf{LLMs for SR.}
LLMs were first applied to recommendation as recommenders that generate the next item or predict user preference from a textual history~\cite{p5,tallrec}.
LLM embedders for SR encode items and histories into embeddings, moving the modeling of item transitions from the SR backbone into the encoder and building on LLMs' strength as text embedders~\cite{llmret2}.
Multimodal SR instead leaves these transitions to the backbone, feeding it item features from small pretrained encoders such as BERT~\cite{bert} and CLIP~\cite{clip} that are usually frozen and see one item at a time: MoRec~\cite{mmrec1} replaces item IDs with these features, while M3SRec~\cite{modrec_m3srec} and HM4SR~\cite{modrec_hm4sr} fuse them with ID embeddings through mixture-of-experts layers.
To inject sequence-level CF into the encoder, LLMEmb~\cite{llmemb} distills ID embeddings into an LLM, SLIM~\cite{slim} distills ChatGPT's reasoning over user behavior sequences into a student LLM, and LLM2Rec~\cite{llm2rec} fine-tunes the LLM to predict the next item from interaction sequences before embedding training.
I-LLMRec~\cite{illmrec} extends this line to images by feeding each history item's CLIP feature into a text-only LLM as a token, which the LLM interprets within its text space.

\vspace{+0.1cm}
\noindent\textbf{VLMs for SR.}
VLMs such as Qwen2.5-VL~\cite{qwen2.5-vl}, LLaVA~\cite{llava}, and InternVL3~\cite{internvl3} can read the raw images of an entire history in one context, rather than relying on item-wise CLIP features as multimodal SR and I-LLMRec do.
One line of work uses VLMs as sequential recommenders: MLLM-MSR~\cite{mllmmsr} and HaNoRec~\cite{hanorec} prompt a VLM with a user's history to score or rank candidate next items, so their cost grows with the number of candidates and limits them to small candidate sets.
Another line of work uses VLMs as item encoders: $\text{VLM}_{\text{Prompt}}$~\cite{vlm_prompt} obtains item embeddings by zero-shot prompting a VLM, and NoteLLM-2~\cite{notellm2} fuses visual and textual features at the item level; both embed each item in isolation, without sequence-level CF.
VLM2Rec follows this embedding paradigm but learns sequence-level CF by training the VLM to match each user's multimodal history with the next item.

\vspace{+0.1cm}
\noindent\textbf{Modality Imbalance.}
Gradient modulation methods for multimodal classification, OGM-GE~\cite{modality_balancing_OGM-GE_1} and AGM~\cite{modality_balancing_AGM_2}, rebalance modalities from statistics aggregated over samples, so they cannot adapt to a weaker modality that varies across users' histories.
VLM2Rec instead corrects each user's weaker modality, identified by how well each modality separates the next item from negatives.
Uni-modal auxiliary losses~\cite{modality_gap_gen1} and sample-level modality valuation~\cite{modality_balancing_sample_3} also counter greedy multimodal learning~\cite{modality_gap_gen3}, but for classification.
Relational knowledge distillation~\cite{rkd} and mutual learning~\cite{dml} match similarity distributions across networks, whereas VLM2Rec matches them across the two modalities of one model.

\definecolor{textblue}{HTML}{D1F2EB}
\definecolor{imgorange}{HTML}{FDEBD0}
\newcommand{\texttoken}[1]{%
  \begingroup\setlength{\fboxsep}{1.5pt}\colorbox{textblue}{\vphantom{Ag}#1}\endgroup%
}
\newcommand{\imgtoken}[1]{%
  \begingroup\setlength{\fboxsep}{1.5pt}\colorbox{imgorange}{\vphantom{Ag}#1}\endgroup%
}

\section{Problem Formulation}
\label{sec:pre}

\begin{figure*}[!t]
    \centering
    \includegraphics[width=\textwidth]{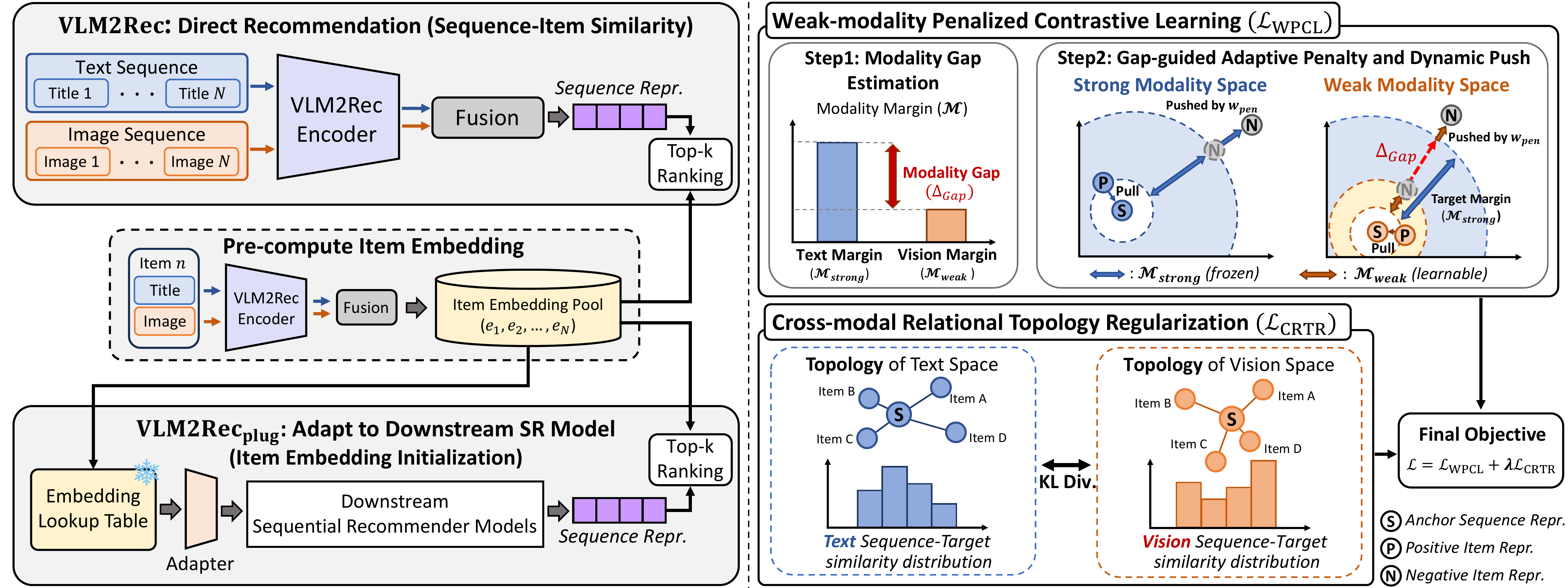}
    \caption{Overview of our framework. Left: VLM2Rec directly ranks items by sequence--item similarity, while $\text{VLM2Rec}_{\text{plug}}$ replaces the item embedding table of a downstream SR backbone with precomputed item embeddings and serves without VLM calls. Right: the shared encoder is trained with $\mathcal{L}_{\text{WPCL}}$ to improve weak-modality discrimination and $\mathcal{L}_{\text{CRTR}}$ to align each user's relational structure across the two modalities.}
    \Description{A unified framework diagram shows direct recommendation with VLM2Rec, downstream SR inference with VLM2Rec plug using precomputed item embeddings, and the shared WPCL and CRTR training objectives.}
    \label{fig:method}
\end{figure*}

Given the set of all items $\mathcal{I}$ and a user's history $S_u=[i_1,\dots,i_L]$, where each item has a title and an image, SR predicts the next item $i^+$.
A VLM $\Phi$ encodes the history into a sequence embedding and each item into an item embedding, taking the $\ell_2$-normalized last-token hidden state of its final layer (Fig.~\ref{fig:method}):
\begin{equation}
\mathbf{e}_u=\Phi(S_u),\qquad \mathbf{e}_i=\Phi(i).
\label{eq:vlm_embed}
\end{equation}
These embeddings serve SR in two ways: ranking all items in $\mathcal{I}$ by the cosine similarity between $\mathbf{e}_u$ and $\mathbf{e}_i$, or replacing the item embedding table of an SR backbone such as SASRec~\cite{sasrec} (Sec.~\ref{sec:deployment}).

\vspace{+0.1cm}
\label{sec:pre_sft}
To turn the generative VLM into an embedder and inject sequence-level CF, we apply standard contrastive Supervised Fine-Tuning (SFT)~\cite{infonce1}.
It pulls each sequence embedding $\mathbf{e}_u$ toward its next-item embedding $\mathbf{e}_{i^+}$ and away from negatives:
\begin{equation}
\mathcal{L}_{\text{SFT}} = - \mathop{\mathbb{E}}\limits_{u \in \mathcal{B}} \log\frac{e^{\text{s}(\mathbf{e}_u, \mathbf{e}_{i^+}) / \tau}}{e^{\text{s}(\mathbf{e}_u, \mathbf{e}_{i^+}) / \tau} + \sum_{i^- \in \mathcal{N}_u} e^{\text{s}(\mathbf{e}_u, \mathbf{e}_{i^-}) / \tau}},
\label{eq:sft_loss}
\end{equation}
where $\mathcal{B}$ is the batch, $\text{s}(\cdot,\cdot)$ is cosine similarity, $\tau$ is the temperature, and $\mathcal{N}_u$ is the negative item set of user $u$.
\section{Analysis of Modality Collapse\\in VLM-based SR}
\label{sec:empirical}

In this section, we analyze SFT-amplified modality collapse in VLM embedders through recommendation performance, gradients, and embedding geometry (Fig.~\ref{fig:drop_grad}, Tab.~\ref{tab:ali_uni_clo}).

\begin{table}[!t]
\centering
\caption{VLM geometry: uniformity $U$ and separability $S$.}
\label{tab:ali_uni_clo}
\setlength{\tabcolsep}{2pt}
\renewcommand{\arraystretch}{0.7}
\resizebox{0.95\columnwidth}{!}{%
\begin{tabular}{@{}lccccccc@{}}
\toprule
\multirow{2}{*}{\textbf{Dataset}} & \multirow{2}{*}{\textbf{Mod.}} & \multicolumn{3}{c}{$U$ ($\downarrow$)} & \multicolumn{3}{c}{$S$ ($\uparrow$)} \\
\cmidrule(lr){3-5}\cmidrule(l){6-8}
 & & $\text{VLM}_{\text{Vanilla}}$ & $\text{VLM}_{\text{SFT}}$ & \textbf{Ours} & $\text{VLM}_{\text{Vanilla}}$ & $\text{VLM}_{\text{SFT}}$ & \textbf{Ours} \\
\midrule
\multirow{3}{*}{\textbf{Beauty}}
 & Fused & -0.3984 & -1.7969 & -1.9921 & 1.0340 & 1.1910 & 1.1870 \\
 & Vision & \textbf{-0.0237} & -0.0398 & \textcolor{blue}{\textbf{-0.2031}} & \textbf{1.0071} & \textcolor{red}{\textbf{1.0000}} & \textcolor{blue}{\textbf{1.2134}} \\
 & Text & -0.3691 & -1.7422 & -1.6718 & 1.0381 & 1.1887 & 1.1639 \\
\midrule
\multirow{3}{*}{\textbf{MovieLens}}
 & Fused & -0.2275 & -0.8203 & -4.6562 & 1.0117 & 1.2471 & 1.2263 \\
 & Vision & \textbf{-0.0236} & \textcolor{red}{\textbf{-0.0078}} & \textcolor{blue}{\textbf{-1.7187}} & \textbf{1.0035} & \textcolor{red}{\textbf{0.9998}} & \textcolor{blue}{\textbf{1.1134}} \\
 & Text & -0.2177 & -0.7891 & -1.1788 & 1.0118 & 1.0116 & 1.1135 \\
\bottomrule
\end{tabular}%
}
\end{table}

\vspace{+0.1cm}
\noindent\textbf{Modality Embeddings.}
\label{sec:prompt_templates}
We encode titles and images in separate prompts, since measuring and correcting each modality requires modality-specific embeddings; each prompt still reads the whole title or image sequence in one context.
Given a user history $S_u$, we build a text prompt $P_T$ and an image prompt $P_V$ and encode each with the same VLM $\Phi$.
\begin{tcolorbox}[colback=gray!5, colframe=gray!50, arc=1mm, boxrule=0.5pt, width=\linewidth, left=6pt, right=6pt, top=3pt, bottom=3pt]
    {\normalsize \textbf{Proposed Separate-Modality Prompts}} \\
    {
    \textbf{[$P_T$]:} Describe this item text sequence in one word, \\
    0) \texttoken{<|item1\_title|>}, \dots, $L{-}1$) \texttoken{<|item$L$\_title|>} \\[2pt]
    \textbf{[$P_V$]:} Describe this item image sequence in one word, \\
    0) \imgtoken{<|item1\_image|>}, \dots, $L{-}1$) \imgtoken{<|item$L$\_image|>}
    }
\end{tcolorbox}
\noindent The instruction ``in one word'' encourages the last token, whose hidden state becomes the embedding, to summarize the entire title (or image) sequence.
We take $\mathbf{e}_u^m=\Phi(P_m(S_u^m))$ and $\mathbf{e}_i^m=\Phi(P_m(i^m))$ for $m\in\{T,V\}$, where $S_u^T$ and $S_u^V$ are the title and image sequences of $S_u$ and $i^m$ is the title or image of item $i$.
Item embeddings use the same prompt with a single item ($L{=}1$).
We then fuse the two modalities by summation without extra parameters: $\mathbf{e}_u=\mathbf{e}_u^T+\mathbf{e}_u^V$ and $\mathbf{e}_i=\mathbf{e}_i^T+\mathbf{e}_i^V$.

\vspace{+0.1cm}
\noindent\textbf{A1: Adding Vision Can Hurt.}
We first examine how standard SFT (Eq.~\ref{eq:sft_loss}) affects modality-specific recommendation performance.
In Fig.~\ref{fig:drop_grad} left, \TEXTset{}, \VISIONset{}, and \FUSIONset{} rank items by the similarity of the text-only ($\mathbf{e}_u^T$), vision-only ($\mathbf{e}_u^V$), and fused ($\mathbf{e}_u$) sequence embeddings to the fused item embeddings $\mathbf{e}_i$.
The Vanilla model ($\text{VLM}_\text{Vanilla}$) already shows a large gap of \TEXTset{} over \VISIONset{} (6.9\% and 7.2\%), reflecting the inherent text-dominant prior of VLMs; yet visual signals can still add complementary preference evidence on some domains (\FUSIONset{}<\TEXTset{} on Beauty but \FUSIONset{}>\TEXTset{} on MovieLens).
However, $\text{VLM}_\text{SFT}$, though necessary to adapt VLMs to SR, worsens this collapse: the \TEXTset{}--\VISIONset{} gap widens to 12.1\% and 14.6\%, and \FUSIONset{} ends up below \TEXTset{} on both datasets. 
Adding visual embeddings now causes negative transfer, lowering next-item accuracy rather than improving it.
To rule out LoRA updates as the cause, we freeze the VLM and train only a linear head with the same loss, and the gap still widens (App.~\ref{app:head_sft_control}).

\vspace{+0.1cm}
\noindent\textbf{A2: SFT Widens the Text--Vision Gradient Gap.}
To examine this degradation, we feed the fused, text-only, and vision-only sequence embeddings into the SFT loss (Eq.~\ref{eq:sft_loss}) and take the gradient with respect to each, denoted by $\mathbf{g}_{\text{fusion}}$, $\mathbf{g}_T$, and $\mathbf{g}_V$. We then measure their alignment $\cos(\mathbf{g}_{\text{fusion}},\mathbf{g}_m)$ for $m\in\{T,V\}$ (Fig.~\ref{fig:drop_grad} right).
This alignment shows how closely the fused gradient follows each modality's own gradient in direction, and we call the difference between the two the \textit{gradient gap} (shaded in Fig.~\ref{fig:drop_grad} right).
From the start, $\mathbf{g}_{\text{fusion}}$ is strongly aligned with the text gradient $\mathbf{g}_T$, while alignment with the vision gradient $\mathbf{g}_V$ drops rapidly.
As SFT pulls each user's sequence embedding toward the next item, $\cos(\mathbf{g}_{\text{fusion}},\mathbf{g}_T)$ rises, whereas $\cos(\mathbf{g}_{\text{fusion}},\mathbf{g}_V)$ only partly recovers and then declines as the imbalance accumulates.
Under SFT, the gradient gap stays large, averaging 0.74 on Beauty and 0.87 on MovieLens over the last 20\% of training (smoothed curves), whereas VLM2Rec closes it ($-0.27$ and $-0.08$).
The large gap under SFT is consistent with the model learning sequence-level CF mainly through the easier text modality as a \textit{shortcut}~\cite{shortcut2}, although these embedding-level gradients do not directly measure each modality's parameter update.

\vspace{+0.1cm}
\noindent\textbf{A3: SFT Sharpens the Text Space, Not the Visual Space.}
To see how this bias shapes the sequence--item representation space used to rank next items, we measure uniformity $U^m$ and alignment $A^m$~\cite{align} between sequence and item embeddings (Tab.~\ref{tab:ali_uni_clo}).
$U^m$ measures how evenly the embeddings spread over the unit hypersphere and approaches 0 as the space collapses, while $A^m_{\text{pos}}$ and $A^m_{\text{neg}}$ measure the squared distance from a user's sequence embedding to the next item and to negatives, respectively:
\begin{equation}
\begin{gathered}
U^{m}
\triangleq
\log
\mathbb{E}_{(u,i)\sim \mathcal{P}_{\text{uniform}}}
\Big[e^{-2\|\mathbf{e}_u^m-\mathbf{e}_i^m\|_2^2}\Big],\\
A_{\text{pos}}^{m} \triangleq
\mathbb{E}_{(u,i^+)\sim \mathcal{D}}
\Big[\|\mathbf{e}_u^m-\mathbf{e}_{i^+}^m\|_2^2\Big],\\
A_{\text{neg}}^{m}
\triangleq
\mathbb{E}_{(u,i^+)\sim \mathcal{D}}
\mathbb{E}_{i^- \sim \mathcal{P}_n(\cdot \mid u)}
\Big[\|\mathbf{e}_u^m-\mathbf{e}_{i^-}^m\|_2^2\Big],
\end{gathered}
\end{equation}
where $\mathcal{D}$ holds evaluation user--next-item pairs, $\mathcal{P}_n(\cdot\mid u)$ samples $u$'s negatives, and $\mathcal{P}_{\text{uniform}}$ samples random sequence--item pairs.
Since ranking depends on whether the next item lies closer to the user's sequence than negatives do, not on absolute alignment, we derive separability $S^m$ from the alignment terms:
\begin{equation}
S^m \triangleq A^m_{\text{neg}} / (A^m_{\text{pos}} + \epsilon),
\end{equation}
where $\epsilon$ is a small constant; $S^m>1$ means that, on average, the modality places users' next items closer than negatives, and $S^m$ near 1 means that it barely does so.
As shown in Tab.~\ref{tab:ali_uni_clo}, under SFT the fused space is closer to the text space than to the vision space in both $U^m$ and $S^m$, indicating that text dominates the fused embeddings.
This is consistent with A2: SFT injects sequence-level CF mainly through text, improving the text space and, in turn, the fused space.
However, the visual modality receives little of this CF signal: its separability stays near 1, far below that of text, and its uniformity lags behind text, indicating collapse in the visual sequence--item space.

\vspace{+0.1cm}
\noindent\textbf{Summary.}
Overall, SFT amplifies the modality collapse that is already present in VLMs.
Its fused gradient mostly follows the text gradient, and the weak visual modality's representation space collapses further, which widens the text--vision accuracy gap and causes negative transfer when fused.
The same pattern appears on Toys and the visually informative Clothing domain (App.~\ref{app:additional_analysis}), and SFT leaves a sizable \TEXTset{}--\VISIONset{} gap across VLM families and scales (Tab.~\ref{tab:modality_collapse_backbones}).

\section{Proposed Method: VLM2Rec}
\label{sec:methods}

To address the SFT-amplified modality collapse (Sec.~\ref{sec:empirical}), we propose \textbf{VLM2Rec} (Fig.~\ref{fig:method}), which trains a VLM with two objectives, without changing its architecture, to recommend the next item by sequence--item similarity.
We follow the modality embedding setting of Sec.~\ref{sec:prompt_templates}.
$\mathcal{L}_{\text{WPCL}}$ strengthens each user's weaker modality (Sec.~\ref{sec:wpcl}), and $\mathcal{L}_{\text{CRTR}}$ softly aligns each user's relative similarity structure over the same items in the text and vision spaces (Sec.~\ref{sec:crtr}); the final objective combines them (Sec.~\ref{sec:final_objective}).
We also propose $\text{VLM2Rec}_{\text{plug}}$ for efficient serving; it transfers the trained item embeddings to SR backbones such as SASRec~\cite{sasrec}, so its latency matches that of conventional SR models (Sec.~\ref{sec:deployment}).

\subsection{Weak-modality Penalized Contrastive Learning}
\label{sec:wpcl}
We propose \textit{Weak-modality Penalized Contrastive Learning} ($\mathcal{L}_{\text{WPCL}}$), which multiplies each user's negative term in $\mathcal{L}_{\text{SFT}}$ (Eq.~\ref{eq:sft_loss}) by a penalty weight that grows as the user's weaker modality lags behind the stronger one.
$\mathcal{L}_{\text{WPCL}}$ targets the imbalance in Sec.~\ref{sec:empirical}: SFT reduces the loss mainly through the strong modality as a shortcut (A2), so the weak modality barely separates next items from negatives (A3).

\subsubsection{\textbf{User-adaptive Modality Gap Estimation}}
We first measure how clearly each modality separates a user's next item from negatives, as a per-user counterpart of the separability $S^m$ (Sec.~\ref{sec:empirical}).
For user $u$ and modality $m\in\{T, V\}$, the \textit{Discriminative Margin} $\mathcal{M}_{u,m}$ is the similarity of the sequence embedding $\mathbf{e}_u^m$ to the next item minus the mean similarity to the negatives $i^-\in\mathcal{N}_u$:
\begin{equation}
    \mathcal{M}_{u,m} = \text{s}(\mathbf{e}^m_u, \mathbf{e}^m_{i^+}) - \frac{1}{|\mathcal{N}_u|} \sum_{i^- \in \mathcal{N}_u} \text{s}(\mathbf{e}^m_u, \mathbf{e}^m_{i^-}),
\end{equation}
where $\mathcal{N}_u$ holds the next items of the other users in the batch (Sec.~\ref{sec:exp_setup}).
At each training step, user $u$'s strong and weak modalities are those with the larger and smaller margins:
\begin{equation}
    \mathcal{M}_{u,\text{strong}} = \max_{m}(\mathcal{M}_{u,m}), \quad \mathcal{M}_{u,\text{weak}} = \min_{m}(\mathcal{M}_{u,m}).
\end{equation}
We identify the weak modality per user because it varies across users (App.~\ref{app:anchor_weak_split}).
$\mathcal{L}_{\text{WPCL}}$ can thus strengthen text for some users and vision for others, unlike gradient modulation methods~\cite{modality_balancing_OGM-GE_1, modality_balancing_AGM_2} based on statistics aggregated over samples (Sec.~\ref{sec:related}).

We define the \textit{Modality Gap} $\Delta_{u,\mathrm{gap}}$ as the difference between these margins:
\begin{equation}
    \Delta_{u,\mathrm{gap}} = \mathrm{sg}[\mathcal{M}_{u,\text{strong}}] - \mathcal{M}_{u,\text{weak}},
    \label{eq:modality_gap}
\end{equation}
where $\mathrm{sg}[\cdot]$ is the stop-gradient operator.
It treats $\mathcal{M}_{u,\text{strong}}$ as a fixed target inside $\Delta_{u,\mathrm{gap}}$ only, so the penalty below can shrink the gap by raising $\mathcal{M}_{u,\text{weak}}$, not by lowering $\mathcal{M}_{u,\text{strong}}$.

\subsubsection{\textbf{Dynamic Penalty from Modality Gap}}
We convert $\Delta_{u,\mathrm{gap}}$ into a penalty weight $w_{u,\text{pen}}$:
\begin{equation}
\label{eq:wpen}
w_{u,\text{pen}}
=
1+\mathrm{Softplus}(\Delta_{u,\mathrm{gap}}),
\end{equation}
Since Softplus is positive and increasing, $w_{u,\text{pen}}>1$ and increases monotonically with $\Delta_{u,\mathrm{gap}}$.
Multiplying each user's negative term in Eq.~\ref{eq:sft_loss} by $w_{u,\text{pen}}$ gives
\begin{equation}
\label{eq:wpcl}
    \mathcal{L}_{\text{WPCL}} = -\mathop{\mathbb{E}}\limits_{u \in \mathcal{B}} \log \frac{e^{\text{s}(\mathbf{e}_u, \mathbf{e}_{i^+}) / \tau_{\text{WPCL}}}}{e^{\text{s}(\mathbf{e}_u, \mathbf{e}_{i^+}) / \tau_{\text{WPCL}}} + w_{u,\text{pen}} \sum_{i^- \in \mathcal{N}_u} e^{\text{s}(\mathbf{e}_u, \mathbf{e}_{i^-}) / \tau_{\text{WPCL}}}},
\end{equation}
\noindent where $\tau_{\text{WPCL}}$ is the temperature.
Minimizing Eq.~\ref{eq:wpcl} also shrinks $\Delta_{u,\mathrm{gap}}$ by raising $\mathcal{M}_{u,\text{weak}}$, because $w_{u,\text{pen}}$ is not detached.

\subsection{Cross-modal Relational Topology Regularization}
\label{sec:crtr}

While $\mathcal{L}_{\text{WPCL}}$ strengthens the weak modality's discrimination, it does not relate the two modalities' similarity structures.
To satisfy the cross-modal consistency that multimodal SR also requires, we introduce \textit{Cross-modal Relational Topology Regularization} ($\mathcal{L}_{\text{CRTR}}$).
It softly aligns each user's relative similarity structure over the same items in the two modality spaces, so that fusing them improves recommendation.
Unlike direct alignment of $\mathbf{e}_u^T$ with $\mathbf{e}_u^V$, it does not require identical text and vision embeddings (Fig.~\ref{fig:method}).
$\mathcal{L}_{\text{WPCL}}$, which scores the summed embeddings, keeps the two spaces compatible for fusion, since $\mathcal{L}_{\text{CRTR}}$ ignores how they are oriented.

\subsubsection{\textbf{Modality-specific Relational Distribution}}
For user $u$, we define the modality-specific relational distribution $\mathbf{P}^m_u \in \mathbb{R}^{|\mathcal{C}|}$ over the candidate items $\mathcal{C}$, where each entry $\mathbf{P}_{u,j}^m$ is the softmax-normalized similarity between the sequence representation $\mathbf{e}_u^m$ of user $u$ and the candidate item embedding $\mathbf{e}_j^m$ of item $j$:
\begin{equation}
    \mathbf{P}_{u,j}^m = \frac{e^{\text{s}(\mathbf{e}_u^m, \mathbf{e}_j^m) / \tau_{\text{CRTR}}}}{\sum_{k\in\mathcal{C}} e^{\text{s}(\mathbf{e}_u^m, \mathbf{e}_k^m) / \tau_{\text{CRTR}}}},
\end{equation}
where $\tau_{\text{CRTR}}$ is a temperature parameter.
$\mathbf{P}_u^m$ captures the neighborhood of user $u$'s history among the candidate items in the modality-specific representation space.

\subsubsection{\textbf{Bidirectional Relational Consistency}}
For each user $u$, we then align the two neighborhoods $\mathbf{P}^T_u$ and $\mathbf{P}^V_u$ by minimizing their bidirectional Kullback--Leibler (KL) divergence:
\begin{equation}
\mathcal{L}_{\text{CRTR}} = \frac{1}{2}\mathop{\mathbb{E}}\limits_{u \in \mathcal{B}} \left( \mathrm{KL}(\mathbf{P}_u^T \| \mathbf{P}_u^V) + \mathrm{KL}(\mathbf{P}_u^V \| \mathbf{P}_u^T) \right).
\end{equation}
The bidirectional form pulls each modality's neighborhood toward the other's, rather than using the dominant modality as a fixed target.




\subsection{Final Objective}
\label{sec:final_objective}
The final objective function $\mathcal{L}$ is defined as follows:
\begin{equation}
\mathcal{L} = \mathcal{L}_{\text{WPCL}} + \lambda \cdot \mathcal{L}_{\text{CRTR}},
\end{equation}
where $\lambda$ balances discrimination and cross-modal consistency.
After training, VLM2Rec encodes a user's history into the fused sequence embedding $\mathbf{e}_u$ and directly ranks items by $\text{s}(\mathbf{e}_u,\mathbf{e}_i)$.

\begin{algorithm}[!h]
\caption{Training and inference of VLM2Rec and $\text{VLM2Rec}_{\text{plug}}$}
\label{alg:vlm2rec_pipeline}
\small
\begin{algorithmic}[1]
\Require Sequence--target pairs $(S_u,i^+)$, item catalog $\mathcal I$, VLM $\Phi$
\For{each minibatch $\mathcal B$ during training}
    \State Encode histories and targets with $P_T,P_V$; normalize last-token states and sum modalities.
    \State Compute per-modality margins; identify weak/strong, then $\Delta_{u,\mathrm{gap}}$ and $w_{u,\text{pen}}$ (Eqs.~\ref{eq:modality_gap} and~\ref{eq:wpen}).
    \State Compute $\mathcal L_{\text{WPCL}}$ (Eq.~\ref{eq:wpcl}) and $\mathcal L_{\text{CRTR}}$ from $\mathbf P_u^T$ and $\mathbf P_u^V$.
    \State Update VLM LoRA with $\mathcal L_{\text{WPCL}}+\lambda\mathcal L_{\text{CRTR}}$.
\EndFor
\State \textbf{Precompute:} Encode each item once into the fused item table $\mathbf E$.
\State \textbf{VLM2Rec:} At each request, encode the history with $\Phi$; rank items by $\text{s}(\mathbf e_u,\mathbf e_i)$ over $\mathbf E$.
\State \textbf{$\text{VLM2Rec}_{\text{plug}}$ (offline):} Replace the item embedding table of an SR backbone $f_\psi$ with the frozen $\mathbf E$ through $\mathbf W$; train only $\mathbf W$ and $\psi$.
\State \textbf{$\text{VLM2Rec}_{\text{plug}}$ (online):} Rank items with the trained SR backbone on $\mathbf E$, without VLM calls.
\end{algorithmic}
\end{algorithm}

\setlength{\fboxsep}{1pt}

\begin{table*}[!t]
\centering
\caption{Performance of VLM2Rec and $\text{VLM2Rec}_{\text{plug}}$. The best and second-best results within each setting are highlighted by \protect\colorbox{bestcolor}{\textbf{boldface}} and \protect\colorbox{secondcolor}{\underline{underline}}. * indicates statistical significance against the strongest baseline ($p < 0.05$, paired t-test over 5 seeds).}
\setlength{\tabcolsep}{1.5pt}
\resizebox{\linewidth}{!}{
\renewcommand{\arraystretch}{0.8}
\begin{tabular}{c|cccc|cccc|cccc|cccc|cccc}
\toprule
\multirow{2.5}{*}{\textbf{Method}} & \multicolumn{4}{c|}{\textbf{Toys}} & \multicolumn{4}{c|}{\textbf{Beauty}} & \multicolumn{4}{c|}{\textbf{Clothing}} & \multicolumn{4}{c|}{\textbf{Sports}} & \multicolumn{4}{c}{\textbf{MovieLens}} \\
\cmidrule(lr){2-5} \cmidrule(lr){6-9} \cmidrule(lr){10-13} \cmidrule(lr){14-17} \cmidrule(lr){18-21}
 & H@10 & H@20 & N@10 & N@20 & H@10 & H@20 & N@10 & N@20 & H@10 & H@20 & N@10 & N@20 & H@10 & H@20 & N@10 & N@20 & H@10 & H@20 & N@10 & N@20 \\
\midrule[\heavyrulewidth]
GRU4Rec & 0.0662 & 0.0963 & 0.0362 & 0.0436 & 0.0638 & 0.0955 & 0.0350 & 0.0428 & 0.0312 & 0.0519 & 0.0153 & 0.0203 & 0.0326 & 0.0504 & 0.0171 & 0.0214 & 0.0962 & 0.1495 & 0.0471 & 0.0593 \\
FMLP & 0.0840 & 0.1138 & 0.0487 & 0.0561 & 0.0665 & 0.0987 & 0.0369 & 0.0451 & 0.0349 & 0.0564 & 0.0173 & 0.0224 & 0.0375 & 0.0564 & 0.0208 & 0.0255 & 0.1091 & 0.1602 & 0.0547 & 0.0673 \\
DiffuRec & 0.0890 & 0.1240 & 0.0500 & 0.0590 & 0.0851 & 0.1225 & 0.0471 & 0.0568 & 0.0488 & 0.0730 & 0.0252 & 0.0315 & 0.0473 & 0.0696 & 0.0250 & 0.0309 & 0.1296 & 0.1901 & 0.0643 & 0.0787 \\
SASRec & 0.0904 & 0.1221 & 0.0524 & 0.0602 & 0.0713 & 0.1053 & 0.0397 & 0.0483 & 0.0402 & 0.0648 & 0.0198 & 0.0260 & 0.0428 & 0.0648 & 0.0240 & 0.0294 & 0.1263 & 0.1830 & 0.0629 & 0.0773 \\
\midrule[\heavyrulewidth]
\multicolumn{21}{c}{\cellcolor{gray!30}\textbf{VLM2Rec Setting (without SR Backbone)}} \\ \midrule
BERT & 0.0550 & 0.0810 & 0.0310 & 0.0376 & 0.0433 & 0.0686 & 0.0223 & 0.0286 & 0.0234 & 0.0408 & 0.0113 & 0.0155 & 0.0244 & 0.0398 & 0.0119 & 0.0157 & 0.0679 & 0.1036 & 0.0320 & 0.0408 \\
SLIM & 0.0576 & 0.0850 & 0.0323 & 0.0390 & 0.0443 & 0.0708 & 0.0232 & 0.0297 & 0.0241 & 0.0426 & 0.0118 & 0.0162 & 0.0254 & 0.0412 & 0.0124 & 0.0164 & 0.0695 & 0.1066 & 0.0327 & 0.0420 \\
SLIM$^{+}$ & 0.0971 & 0.1379 & 0.0547 & 0.0647 & 0.0795 & 0.1213 & 0.0439 & 0.0544 & 0.0430 & 0.0673 & 0.0215 & 0.0276 & 0.0419 & 0.0640 & 0.0220 & 0.0276 & 0.1270 & 0.1909 & 0.0618 & 0.0780 \\
$\text{LLM}_{\text{Vanilla}}$ & 0.0781 & 0.1095 & 0.0436 & 0.0516 & 0.0593 & 0.0913 & 0.0342 & 0.0422 & 0.0332 & 0.0545 & 0.0169 & 0.0222 & 0.0351 & 0.0563 & 0.0192 & 0.0244 & 0.1133 & 0.1739 & 0.0547 & 0.0699 \\
$\text{LLM}_{\text{SFT}}$ & 0.1041 & 0.1467 & 0.0586 & 0.0693 & 0.0952 & 0.1539 & 0.0485 & 0.0631 & 0.0489 & 0.0745 & 0.0242 & 0.0306 & 0.0511 & 0.0756 & 0.0266 & 0.0327 & 0.1331 & 0.1975 & 0.0656 & 0.0817 \\
\midrule
$\text{VLM}_{\text{Vanilla}}$ & 0.0843 & 0.1197 & 0.0481 & 0.0570 & 0.0674 & 0.1054 & 0.0351 & 0.0442 & 0.0384 & 0.0610 & 0.0189 & 0.0245 & 0.0415 & 0.0605 & 0.0224 & 0.0271 & 0.1188 & 0.1740 & 0.0570 & 0.0711 \\
$\text{VLM}_{\text{SFT}}$ & \secondscore{0.1132} & 0.1456 & \secondscore{0.0652} & \secondscore{0.0732} & 0.0942 & 0.1527 & 0.0547 & 0.0695 & 0.0471 & \secondscore{0.0807} & 0.0241 & 0.0325 & 0.0506 & \secondscore{0.0817} & 0.0279 & \secondscore{0.0358} & 0.1671 & \secondscore{0.2527} & 0.0869 & \secondscore{0.1087} \\
MLLM-MSR & 0.1092 & 0.1504 & 0.0608 & 0.0711 & 0.0994 & 0.1573 & 0.0512 & 0.0655 & 0.0491 & 0.0750 & 0.0248 & 0.0313 & 0.0521 & 0.0788 & 0.0273 & 0.0342 & 0.1552 & 0.2259 & 0.0805 & 0.0982 \\
I-LLMRec & 0.1116 & \secondscore{0.1549} & 0.0621 & 0.0730 & \secondscore{0.1047} & \secondscore{0.1629} & \secondscore{0.0554} & \secondscore{0.0700} & \secondscore{0.0517} & 0.0803 & \secondscore{0.0258} & \secondscore{0.0331} & \secondscore{0.0545} & 0.0808 & \secondscore{0.0292} & 0.0357 & \secondscore{0.1699} & 0.2474 & \secondscore{0.0886} & 0.1085 \\
\midrule[\heavyrulewidth]
\textbf{VLM2Rec} & \bestscore{0.1254*} & \bestscore{0.1682*} & \bestscore{0.0720*} & \bestscore{0.0823*} & \bestscore{0.1161*} & \bestscore{0.1732*} & \bestscore{0.0641*} & \bestscore{0.0784*} & \bestscore{0.0553*} & \bestscore{0.0875*} & \bestscore{0.0290*} & \bestscore{0.0370*} & \bestscore{0.0585*} & \bestscore{0.0904*} & \bestscore{0.0316*} & \bestscore{0.0396*} & \bestscore{0.1899*} & \bestscore{0.2783*} & \bestscore{0.1029*} & \bestscore{0.1252*} \\
\textit{Imp. (\%)} & 10.78 & 8.59 & 10.43 & 12.43 & 10.89 & 6.32 & 15.70 & 12.00 & 6.96 & 8.43 & 12.40 & 11.78 & 7.34 & 10.65 & 8.22 & 10.61 & 11.77 & 10.13 & 16.14 & 15.18 \\
\midrule[\heavyrulewidth]
\multicolumn{21}{c}{\cellcolor{gray!30}\textbf{$\text{VLM2Rec}_{\text{plug}}$ Setting (Backbone: SASRec)}} \\ \midrule
BERT & 0.0702 & 0.1037 & 0.0387 & 0.0471 & 0.0601 & 0.0952 & 0.0308 & 0.0395 & 0.0303 & 0.0517 & 0.0144 & 0.0197 & 0.0314 & 0.0511 & 0.0157 & 0.0207 & 0.0884 & 0.1337 & 0.0397 & 0.0510 \\
SLIM & 0.0733 & 0.1077 & 0.0412 & 0.0497 & 0.0606 & 0.0957 & 0.0311 & 0.0397 & 0.0306 & 0.0525 & 0.0147 & 0.0201 & 0.0324 & 0.0519 & 0.0162 & 0.0210 & 0.0901 & 0.1347 & 0.0403 & 0.0512 \\
SLIM$^{+}$ & 0.0961 & 0.1332 & 0.0561 & 0.0655 & 0.0812 & 0.1205 & 0.0443 & 0.0540 & 0.0434 & 0.0681 & 0.0218 & 0.0280 & 0.0427 & 0.0656 & 0.0229 & 0.0287 & 0.1555 & 0.2341 & 0.0779 & 0.0978 \\
LLMEmb & 0.0985 & 0.1371 & 0.0574 & 0.0667 & 0.0854 & 0.1247 & 0.0468 & 0.0564 & 0.0452 & 0.0692 & 0.0234 & 0.0292 & 0.0442 & 0.0680 & 0.0241 & 0.0299 & 0.1588 & 0.2356 & 0.0796 & 0.0985 \\
LLM2Rec & 0.1004 & 0.1403 & 0.0592 & 0.0691 & 0.0896 & 0.1299 & 0.0493 & 0.0594 & 0.0470 & 0.0718 & 0.0244 & 0.0305 & 0.0461 & 0.0691 & 0.0253 & 0.0312 & 0.1598 & 0.2367 & 0.0827 & 0.1018 \\
$\text{LLM}_{\text{Vanilla}}$ & 0.0816 & 0.1142 & 0.0457 & 0.0540 & 0.0645 & 0.0985 & 0.0337 & 0.0425 & 0.0341 & 0.0563 & 0.0164 & 0.0220 & 0.0367 & 0.0575 & 0.0199 & 0.0250 & 0.1095 & 0.1654 & 0.0538 & 0.0678 \\
$\text{LLM}_{\text{SFT}}$ & 0.1015 & 0.1407 & 0.0616 & 0.0715 & 0.0948 & 0.1365 & 0.0506 & 0.0612 & 0.0479 & 0.0724 & 0.0249 & 0.0310 & 0.0492 & 0.0716 & 0.0268 & 0.0325 & 0.1624 & 0.2374 & 0.0865 & 0.1052 \\
\midrule
CLIP & 0.0735 & 0.1044 & 0.0408 & 0.0484 & 0.0663 & 0.0997 & 0.0358 & 0.0440 & 0.0344 & 0.0578 & 0.0173 & 0.0233 & 0.0357 & 0.0563 & 0.0191 & 0.0243 & 0.1033 & 0.1560 & 0.0506 & 0.0636 \\
$\text{VLM}_{\text{Prompt}}$ & 0.0792 & 0.1123 & 0.0449 & 0.0530 & 0.0679 & 0.1008 & 0.0367 & 0.0449 & 0.0359 & 0.0594 & 0.0178 & 0.0235 & 0.0374 & 0.0588 & 0.0204 & 0.0256 & 0.1115 & 0.1668 & 0.0548 & 0.0690 \\
NoteLLM-2 & 0.1049 & 0.1438 & 0.0596 & 0.0690 & 0.0903 & 0.1309 & 0.0507 & 0.0608 & 0.0492 & 0.0744 & 0.0258 & 0.0320 & 0.0470 & 0.0730 & 0.0261 & 0.0324 & 0.1631 & 0.2396 & 0.0885 & 0.1072 \\
$\text{VLM}_{\text{Vanilla}}$ & 0.0841 & 0.1178 & 0.0491 & 0.0577 & 0.0685 & 0.1031 & 0.0374 & 0.0463 & 0.0365 & 0.0596 & 0.0183 & 0.0240 & 0.0383 & 0.0601 & 0.0208 & 0.0262 & 0.1152 & 0.1695 & 0.0559 & 0.0695 \\
$\text{VLM}_{\text{SFT}}$ & 0.1058 & \secondscore{0.1461} & 0.0598 & 0.0699 & 0.0922 & 0.1349 & 0.0515 & 0.0622 & \secondscore{0.0499} & 0.0755 & \secondscore{0.0260} & 0.0325 & 0.0469 & \secondscore{0.0738} & 0.0260 & 0.0326 & \secondscore{0.1649} & 0.2365 & \secondscore{0.0902} & 0.1082 \\
MLLM-MSR & 0.1038 & 0.1421 & 0.0611 & 0.0706 & 0.0940 & 0.1340 & 0.0506 & 0.0604 & 0.0482 & 0.0730 & 0.0253 & 0.0316 & 0.0494 & 0.0725 & 0.0272 & 0.0329 & 0.1627 & 0.2402 & 0.0864 & 0.1057 \\
I-LLMRec & \secondscore{0.1064} & 0.1442 & \secondscore{0.0624} & \secondscore{0.0717} & \secondscore{0.0967} & \secondscore{0.1390} & \secondscore{0.0539} & \secondscore{0.0644} & 0.0483 & \secondscore{0.0759} & 0.0258 & \secondscore{0.0328} & \secondscore{0.0503} & 0.0737 & \secondscore{0.0278} & \secondscore{0.0337} & 0.1644 & \secondscore{0.2461} & 0.0881 & \secondscore{0.1085} \\
\midrule[\heavyrulewidth]
\textbf{$\text{VLM2Rec}_{\text{plug}}$} & \bestscore{0.1169*} & \bestscore{0.1588*} & \bestscore{0.0684*} & \bestscore{0.0786*} & \bestscore{0.1080*} & \bestscore{0.1511*} & \bestscore{0.0618*} & \bestscore{0.0728*} & \bestscore{0.0540*} & \bestscore{0.0823*} & \bestscore{0.0276*} & \bestscore{0.0345*} & \bestscore{0.0566*} & \bestscore{0.0849*} & \bestscore{0.0307*} & \bestscore{0.0378*} & \bestscore{0.1782*} & \bestscore{0.2718*} & \bestscore{0.0977*} & \bestscore{0.1216*} \\
\textit{Imp. (\%)} & 9.87 & 8.69 & 9.62 & 9.62 & 11.69 & 8.71 & 14.66 & 13.04 & 8.22 & 8.43 & 6.15 & 5.18 & 12.52 & 15.04 & 10.43 & 12.17 & 8.07 & 10.44 & 8.31 & 12.07 \\
\bottomrule
\end{tabular}
}
\label{tab:perf_task2}
\end{table*}

\subsection{$\text{VLM2Rec}_{\text{plug}}$: Plugging into SR Backbones}
\label{sec:deployment}
VLM2Rec runs the VLM on each user's latest history at every request, which is far heavier than a typical SR model.
However, item embeddings depend only on the items and can be computed once offline.
We therefore propose the $\text{VLM2Rec}_{\text{plug}}$ setting, in which we encode every item with the trained VLM2Rec model into a frozen item table $\mathbf{E}=\{\mathbf{e}_i\}_{i\in\mathcal{I}}$ (Fig.~\ref{fig:method}).
$\mathbf{E}$ inherits the sequence-aware knowledge that the model learned by reasoning over multimodal histories, even though each item is encoded alone.
This table then replaces the item embedding table of an SR backbone $f_\psi$ through a linear adapter $\mathbf{W}$:
\begin{equation}
\hat{y}_{u,i}=f_\psi(\mathbf{W}\mathbf{e}_{i_1},\dots,\mathbf{W}\mathbf{e}_{i_L})^{\top}\mathbf{W}\mathbf{e}_i.
\label{eq:sr_score}
\end{equation}
Only $\mathbf{W}$ and $\psi$ are trained with the backbone's loss, so the backbone learns item transitions on embeddings that already carry multimodal semantics and sequence-level CF.
The serving latency matches that of the SR backbone alone, because $\mathbf{E}$ is precomputed and no VLM runs per request.
Algorithm~\ref{alg:vlm2rec_pipeline} summarizes training and inference of both settings.


\section{Experiments}
\label{sec:exp}
\subsection{Experimental Setup}
\label{sec:exp_setup}
\noindent{\textbf{Datasets.}}
We use four Amazon domains~\cite{amazon} (Toys, Beauty, Clothing, Sports) and MovieLens~\cite{ml-1m} to test generalization across domains and platforms on the Web, with 5-core filtering and leave-one-out splits.
Preprocessing and statistics are in App.~\ref{app:datasets} (Tab.~\ref{tab:dataset_stats}).


\vspace{+0.1cm}
\noindent{\textbf{Baselines.}}
The baselines form three groups (details in App.~\ref{app:exp_details}).
\begin{itemize}[wide=0pt]
    \item \textbf{ID-based models}: GRU4Rec~\cite{gru4rec}, FMLP~\cite{fmlp}, DiffuRec~\cite{diffurec}, and SASRec~\cite{sasrec}, which learn item embeddings from interactions alone.
    \item \textbf{Text-based models}: BERT~\cite{bert}, a small text encoder; SLIM and SLIM$^{+}$~\cite{slim}, which embed items with BERT and its CF-trained student LLM, respectively; LLMEmb~\cite{llmemb} and LLM2Rec~\cite{llm2rec}, which inject sequence-level CF into LLM embeddings; and $\text{LLM}_{\text{Vanilla}}$ and $\text{LLM}_{\text{SFT}}$, an untuned LLM and one tuned with standard SFT (Eq.~\ref{eq:sft_loss}).
    \item \textbf{Multimodal-based models}: CLIP~\cite{clip}, a small image--text encoder; $\text{VLM}_{\text{Prompt}}$~\cite{vlm_prompt}, NoteLLM-2~\cite{notellm2}, and I-LLMRec~\cite{illmrec}, which embed items via zero-shot prompting, item-level fusion, and CLIP features fed to an LLM, respectively; MLLM-MSR~\cite{mllmmsr}, which summarizes item images and user preferences into text with a VLM; and $\text{VLM}_{\text{Vanilla}}$ and $\text{VLM}_{\text{SFT}}$, the VLM counterparts of the LLM variants.
\end{itemize}

\vspace{+0.1cm}
\noindent{\textbf{Evaluation Protocol.}} 
We use full-ranking evaluation~\cite{negative_sampling} in two settings.
In the VLM2Rec setting, each method encodes the user's history and ranks all items by $\text{s}(\mathbf{e}_u,\mathbf{e}_i)$.
In the $\text{VLM2Rec}_{\text{plug}}$ setting, a linear adapter (or a baseline's own adapter, when specified) maps each method's frozen item embeddings to the hidden size ($d{=}128$) of SASRec~\cite{sasrec}; only the adapter and the backbone are trained, so all methods share the same inference cost.
We also evaluate DiffuRec~\cite{diffurec}, FMLP~\cite{fmlp}, and GRU4Rec~\cite{gru4rec} as SR backbones (Tab.~\ref{tab:add_backbones}).
We report H@$K$ and N@$K$ ($K \in \{10,20\}$) averaged over five seeds; all gains pass a paired $t$-test ($p<0.05$).

\vspace{+0.1cm}
\noindent{\textbf{Implementation Details.}} We build VLM2Rec and the VLM-based baselines on Qwen2.5-VL-3B and the LLM-based baselines on Qwen2.5-3B, where applicable; tuned variants use LoRA~\cite{lora}.
$\mathcal{N}_u$ and $\mathcal{C}$ are built from in-batch items, and Tab.~\ref{tab:negative_sampling} tests hard negatives for $\mathcal{N}_u$.
Other training details are in App.~\ref{app:implementation_details}.

\subsection{Main Results}
\label{sec:exp_main}
VLM2Rec and $\text{VLM2Rec}_{\text{plug}}$ significantly outperform all baselines (Tab.~\ref{tab:perf_task2}), as three observations explain.

\begin{itemize}[wide=0pt]
    \item \textbf{O1: Semantics alone cannot replace sequence-level CF.}
    SFT improves both Vanilla models by 17--57\% in N@20 across the two settings, and SLIM stays near BERT, whereas $\text{SLIM}^{+}$, which embeds with its CF-trained LLM, improves sharply.
    VLM2Rec keeps the CF that SFT injects, as $\mathcal{L}_{\text{WPCL}}$ retains the SFT objective and only reweights its negative term.
    The encoder needs CF even when the SR backbone learns CF on its own: in the $\text{VLM2Rec}_{\text{plug}}$ setting, encoders trained without CF (BERT, CLIP, $\text{VLM}_{\text{Prompt}}$, and the Vanilla models) all trail even the ID-based SASRec.
    \item \textbf{O2: Correcting the collapse turns vision into consistent gains.}
    Images complement titles: $\text{VLM}_{\text{Vanilla}}$ outperforms $\text{LLM}_{\text{Vanilla}}$ in both settings, as does CLIP over BERT.
    Yet $\text{VLM}_{\text{SFT}}$, though better than $\text{LLM}_{\text{SFT}}$ on most metrics, trails it on several in both settings: SFT-amplified collapse leaves the visual modality under-optimized, causing negative transfer when fused.
    VLM2Rec, which differs from $\text{VLM}_{\text{SFT}}$ only in its objectives, outperforms all baselines, including the multimodal runners-up ($\text{VLM}_{\text{SFT}}$ or I-LLMRec), as does $\text{VLM2Rec}_{\text{plug}}$ in its setting.
    $\mathcal{L}_{\text{WPCL}}$ strengthens each user's weaker modality and $\mathcal{L}_{\text{CRTR}}$ aligns each user's similarity structures across the two modalities, so fusion adds visual evidence instead of causing negative transfer (Sec.~\ref{sec:modality_mute}).
    \item \textbf{O3: The item embeddings carry these gains into SR backbones.}
    Because the $\text{VLM2Rec}_{\text{plug}}$ setting passes only the frozen item embeddings to the backbone, the same trends there show that VLM2Rec's item embeddings themselves carry sequence-level CF (O1) and balanced multimodal semantics (O2).
    Accordingly, $\text{VLM2Rec}_{\text{plug}}$ also performs best with DiffuRec, FMLP, and GRU4Rec (Tab.~\ref{tab:add_backbones}), serving as plug-and-play item representations.
\end{itemize}

\begin{figure}[!t]
    \centering
    \includegraphics[width=\columnwidth]{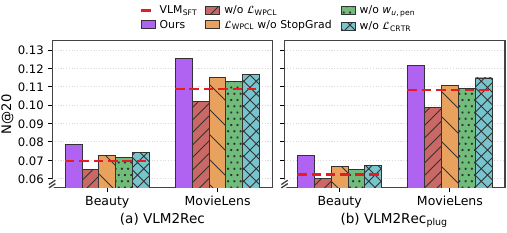}
    \vspace{-0.3cm}
    \caption{Ablation of (a) VLM2Rec and (b) $\text{VLM2Rec}_{\text{plug}}$.}
    \label{fig:ablation_cf}
    \Description{Two bar-chart panels, (a) VLM2Rec and (b) VLM2Rec-plug with SASRec, each with Beauty and MovieLens groups on a shared y-axis. The full model outperforms all four ablations in every group, with the largest decrease when WPCL is removed. Red dashed lines show the VLM SFT baseline.}
\end{figure}

\begin{table}[!t]
\centering
\caption{Generic gradient modulation vs. $\mathcal{L}_{\text{WPCL}}$ in the $\text{VLM2Rec}_{\text{plug}}$ setting (SASRec, N@20).}
\vspace{-0.2cm}
\label{tab:gradient_balancing}
\setlength{\tabcolsep}{5pt}
\renewcommand{\arraystretch}{0.7}
\begin{tabular}{@{}llcccc@{}}
\toprule
\textbf{Dataset} & \textbf{Method} & \textbf{\TEXTset} & \textbf{\VISIONset} & \textbf{\FUSIONset} & \textbf{Gap} (\%) \\
\midrule
\multirow{4}{*}{Beauty}
 & $\text{VLM}_{\text{SFT}}$   & 0.0650 & 0.0587 & 0.0622 & 10.7 \\
 & OGM-GE~\cite{modality_balancing_OGM-GE_1}    & 0.0677 & 0.0616 & 0.0632 & 9.9 \\
 & AGM~\cite{modality_balancing_AGM_2}       & 0.0664 & 0.0605 & 0.0648 & 9.8 \\
 & $\mathcal{L}_{\text{WPCL}}$ only & 0.0667 & 0.0638 & 0.0673 & 4.5 \\
\midrule
\multirow{4}{*}{MovieLens}
 & $\text{VLM}_{\text{SFT}}$   & 0.1100 & 0.0953 & 0.1082 & 15.4 \\
 & OGM-GE~\cite{modality_balancing_OGM-GE_1}    & 0.1087 & 0.0965 & 0.1068 & 12.6 \\
 & AGM~\cite{modality_balancing_AGM_2}       & 0.1120 & 0.1008 & 0.1108 & 11.1 \\
 & $\mathcal{L}_{\text{WPCL}}$ only & 0.1136 & 0.1108 & 0.1148 & 2.5 \\
\bottomrule
\end{tabular}
\end{table}

\subsection{Analysis of Mitigating Modality Collapse}
\label{sec:modality_mute}
VLM2Rec first rebalances optimization: it closes the text--vision gradient gap of $\text{VLM}_{\text{SFT}}$ and even tilts it slightly toward vision, since $\mathcal{L}_{\text{WPCL}}$ adds an extra gradient to each user's weak modality, mostly vision, and $\mathcal{L}_{\text{CRTR}}$ aligns each user's similarity distributions across the two modalities (Fig.~\ref{fig:drop_grad} right).
This restores the visual embedding geometry: visual separability now matches or exceeds that of text, and visual uniformity improves sharply (Tab.~\ref{tab:ali_uni_clo}).
Consequently, the \TEXTset{}--\VISIONset{} gap, which $\text{VLM}_{\text{SFT}}$ widened, nearly closes (1.3\% and 2.0\%), and fusion turns from negative transfer into synergy: \FUSIONset{} > \TEXTset{} and \FUSIONset{} > \VISIONset{} (Fig.~\ref{fig:drop_grad} left).
Across additional domains, model families, and scales, VLM2Rec consistently narrows the \TEXTset{}--\VISIONset{} gap and improves \FUSIONset{} (Fig.~\ref{fig:drop_grad_toys_clothing} and Tab.~\ref{tab:modality_collapse_backbones}).
Even an open-source embedding-pretrained VLM (Qwen3-VL-Embedding) shows a sizable \TEXTset{}--\VISIONset{} gap, which VLM2Rec narrows while improving \FUSIONset{} (Tab.~\ref{tab:qwen3_vl_embedding}).


\begin{table}[!t]

\centering
\caption{Modality embedding replacement experiments on state-of-the-art models. \protect\colorbox{bestcolor}{\textbf{Bold}}/\protect\colorbox{secondcolor}{\underline{underline}}: best/second best.}

\label{tab:modality_rec}
\resizebox{\linewidth}{!}{
    \renewcommand{\arraystretch}{0.7}
    \begin{tabular}{lcccc}
    \toprule
    \multirow{2}{*}{\textbf{Embedding}}
    & \multicolumn{2}{c}{\textbf{UniSRec}~\cite{modrec_unisrec}}
    & \multicolumn{2}{c}{\textbf{TedRec}~\cite{modrec_tedrec}} \\
    \cmidrule(lr){2-3} \cmidrule(lr){4-5}
    & Clothing & MovieLens & Clothing & MovieLens \\
    \midrule
    BERT                              & 0.0308               & 0.1086               & 0.0332               & 0.1167               \\
    $\text{LLM}_{\text{SFT}}$         & 0.0325               & 0.1119               & 0.0348               & 0.1180               \\
    $\text{VLM}_{\text{SFT}}$ (Text)  & \secondscore{0.0328} & 0.1124               & \secondscore{0.0355} & \secondscore{0.1189} \\
    I-LLMRec (Text)                   & 0.0318               & \secondscore{0.1128} & \secondscore{0.0355} & 0.1184               \\
    \textbf{$\text{VLM2Rec}_{\text{plug}}$ (Text)}        & \bestscore{0.0336}   & \bestscore{0.1137}   & \bestscore{0.0362}   & \bestscore{0.1208}   \\
    \midrule
    \multirow{2}{*}{\textbf{Embedding}}
    & \multicolumn{2}{c}{\textbf{M3SRec}~\cite{modrec_m3srec}}
    & \multicolumn{2}{c}{\textbf{HM4SR}~\cite{modrec_hm4sr}} \\
    \cmidrule(lr){2-3} \cmidrule(lr){4-5}
    & Clothing & MovieLens & Clothing & MovieLens \\
    \midrule
    CLIP                      & 0.0322               & 0.1134               & 0.0277               & 0.1219               \\
    $\text{VLM}_{\text{SFT}}$ & 0.0338               & 0.1158               & 0.0328               & \secondscore{0.1261} \\
    I-LLMRec                  & \secondscore{0.0347} & \secondscore{0.1165} & \secondscore{0.0345} & 0.1259               \\
    \textbf{$\text{VLM2Rec}_{\text{plug}}$}       & \bestscore{0.0352}   & \bestscore{0.1176}   & \bestscore{0.0396}   & \bestscore{0.1315}   \\
    \bottomrule
    \end{tabular}
    }

\end{table}

\subsection{Ablation Study}
We ablate each component in both settings to examine its contribution (Fig.~\ref{fig:ablation_cf}).
Every component contributes, with $\mathcal{L}_{\text{WPCL}}$ the most critical, and both settings show the same ordering of variants.
Without $\mathcal{L}_{\text{WPCL}}$, the model underperforms $\text{VLM}_{\text{SFT}}$, as it neither injects sequence-level CF nor improves weak-modality discrimination.
Removing $w_{u,\text{pen}}$ from $\mathcal{L}_{\text{WPCL}}$ hurts performance, showing that the penalty matters.
Without the stop-gradient in $\mathcal{L}_{\text{WPCL}}$, optimization can close $\Delta_{u,\mathrm{gap}}$ by degrading the strong-modality margin rather than improving the weak one, also harming performance.
Removing $\mathcal{L}_{\text{CRTR}}$ lowers performance, suggesting that beyond strengthening the weak modality, fusion benefits from aligning each user's relative similarity structure across the two modality spaces.

\subsection{Analysis of Generic Gradient Modulation}
\label{sec:gradient_balancing}
We compare generic gradient modulation methods, OGM-GE~\cite{modality_balancing_OGM-GE_1} and AGM~\cite{modality_balancing_AGM_2}, with $\mathcal{L}_{\text{WPCL}}$ alone (without $\mathcal{L}_{\text{CRTR}}$) under the same $\text{VLM}_{\text{SFT}}$ setup (Tab.~\ref{tab:gradient_balancing}).
OGM-GE and AGM rescale gradients with modality statistics aggregated over samples, which can ease the average imbalance but cannot adapt to a weaker modality that varies across users.
Instead, $\mathcal{L}_{\text{WPCL}}$ identifies the weaker modality from each user's text and vision margins and adds a $\Delta_{u,\mathrm{gap}}$-dependent discriminative gradient, with the stronger margin held fixed by the stop-gradient.
Consistent with this distinction, the generic methods still show negative transfer (\FUSIONset{} below \TEXTset{}), whereas $\mathcal{L}_{\text{WPCL}}$ more than halves the \TEXTset{}--\VISIONset{} gap and makes fusion outperform either modality.

\subsection{Transfer to Hybrid SR Models}
\label{sec:exp_hybrid_backbones}

To test whether $\text{VLM2Rec}_{\text{plug}}$ remains useful in hybrid recommenders that already have ID embeddings, we replace the modality embeddings of the multimodal M3SRec~\cite{modrec_m3srec} and HM4SR~\cite{modrec_hm4sr} and the text-based UniSRec~\cite{modrec_unisrec} and TedRec~\cite{modrec_tedrec} with those of baseline encoders or $\text{VLM2Rec}_{\text{plug}}$ (Tab.~\ref{tab:modality_rec}).
For the text-based models, the multimodal embedders supply only their text representation $\mathbf{e}_i^T$.
Even the text representation of $\text{VLM}_{\text{SFT}}$ outperforms $\text{LLM}_{\text{SFT}}$, suggesting that visual signals help the VLM learn a better text representation space; $\text{VLM2Rec}_{\text{plug}}$ (Text) improves further, likely because strengthening the weak modality lets the VLM exploit these signals better.
On the multimodal models, $\text{VLM2Rec}_{\text{plug}}$ also performs best, showing that it remains useful alongside ID embeddings.

\begin{table}[!t]
    \centering
    \caption{User-level $\Delta_{u,\mathrm{gap}}$ reduction and its correlation with rank recovery.}
    \label{tab:user_stratification}
    \setlength{\tabcolsep}{3pt}
    \resizebox{\columnwidth}{!}{
    \renewcommand{\arraystretch}{0.8}
    \begin{tabular}{llccc ccc}
    \toprule
     & & \multicolumn{3}{c}{\textbf{Modality gap}} & \multicolumn{3}{c}{\textbf{Pearson $r$}} \\
    \cmidrule(lr){3-5} \cmidrule(lr){6-8}
    \textbf{Data} & \textbf{Group} & $\Delta_{u,\mathrm{gap}}^{\text{SFT}}$ & $\Delta_{u,\mathrm{gap}}^{\text{Ours}}$ & \textbf{User Red.} & \textbf{Weak} & \textbf{Strong} & \textbf{Fused} \\
    \midrule
    \multirow{3}{*}{MovieLens}
    & Low & 0.4048 & 0.0585 & 0.9461 & \textbf{0.16} & 0.01 & 0.04 \\
    & Collapse & 0.6322 & 0.0691 & 1.0000 & \textbf{0.25} & 0.02 & 0.06 \\
    & High & 0.3642 & 0.0679 & 0.9245 & \textbf{0.28} & 0.05 & 0.06 \\
    \midrule
    \multirow{3}{*}{Clothing}
    & Low & 0.1711 & 0.0473 & 0.8410 & \textbf{0.34} & 0.13 & 0.11 \\
    & Collapse & 0.2800 & 0.0561 & 0.9928 & \textbf{0.43} & 0.14 & 0.11 \\
    & High & 0.1696 & 0.0566 & 0.8039 & \textbf{0.46} & 0.11 & 0.06 \\
    \bottomrule
    \end{tabular}
    }
\end{table}

\begin{figure}[!t]
    \centering
    \includegraphics[width=\columnwidth]{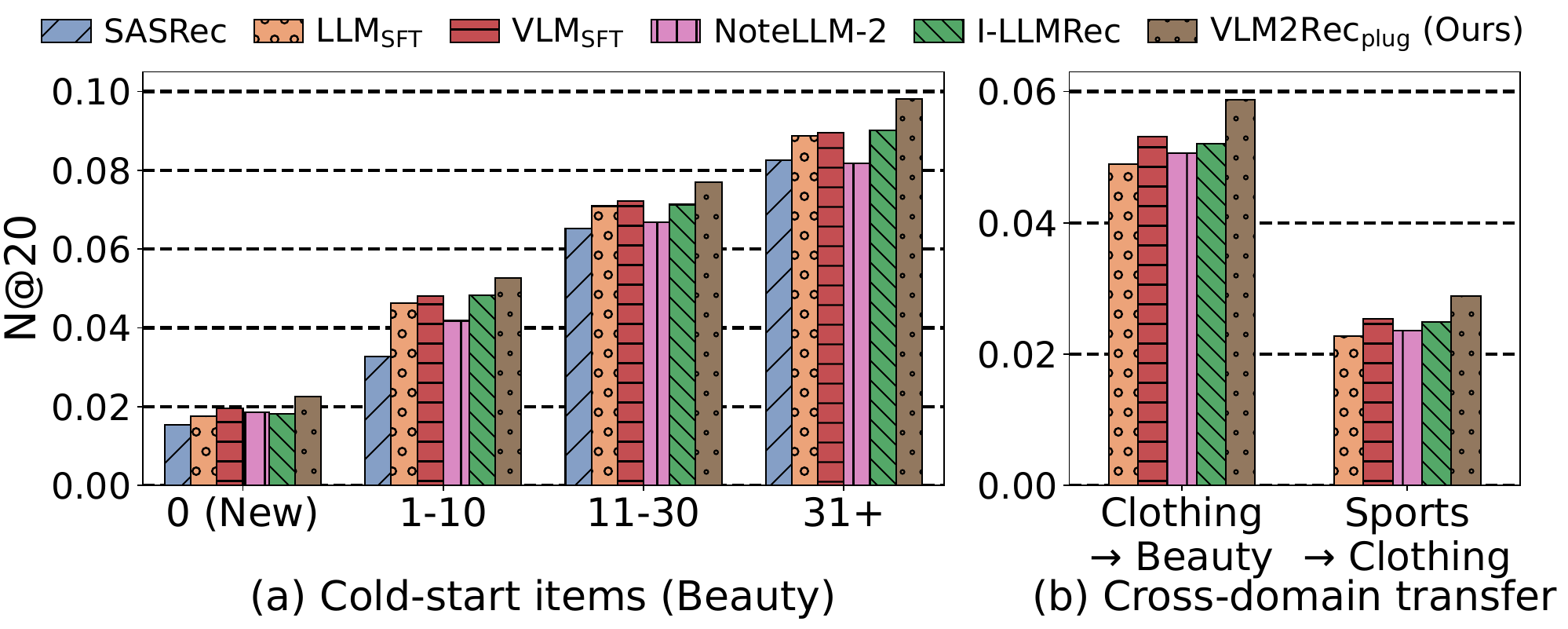}
    \caption{Scarce CF signals: (a) cold-start items, grouped by training interactions; (b) cross-domain transfer.}
    \label{fig:scarce_cf}
\end{figure}

\subsection{User-level Gap and Rank Recovery Analysis}
\label{sec:user_stratification}

We test whether VLM2Rec reduces the gap consistently across users and whether this reduction relates to its ranking gains (Tab.~\ref{tab:user_stratification}).
\textit{Low} and \textit{High} are the users with the bottom and top 20\% of $\Delta_{u,\mathrm{gap}}$ under $\text{VLM}_{\text{Vanilla}}$, and \textit{Collapse} is the half of \textit{Low} users whose gap grows most under SFT.
VLM2Rec lowers the gap from $\Delta_{u,\mathrm{gap}}^{\text{SFT}}$ to $\Delta_{u,\mathrm{gap}}^{\text{Ours}}$ in every group and for most users (User Red., the share of users whose gap shrinks).
Per-user gap recovery correlates more with the weak modality's rank recovery over $\text{VLM}_{\text{SFT}}$ (Pearson $r$ 0.16--0.46) than with the strong or fused one (at most 0.14), so VLM2Rec's gains are tied more to recovering the weak modality than to refining the fused representation.
To check whether each user-level part of the penalty matters, we test three variants of $w_{u,\text{pen}}$ (Tab.~\ref{tab:penalty_variants}).
First, to test whether a fixed margin explains the gain, we compare a \textit{static} penalty $w_{u,\text{pen}}{=}1+\log 2$, its value without a gap, with removing $w_{u,\text{pen}}$: it recovers little of VLM2Rec's gain, so a fixed margin is not its main source.
Second, to test whether the gap must be measured per user, we compare a \textit{batch-level} penalty from batch-averaged margins with the static one: it helps on MovieLens (+5.9\%), where most users are vision-weak, but barely on Clothing (+1.4\%), where text- and vision-weak users nearly split and their gaps largely cancel in the batch average, so the gap must be measured per user.
Third, to test whether the weak modality must be identified per user, we compare a \textit{modality-fixed} penalty that fixes vision as weak with VLM2Rec: it lowers N@20 by 1.9\% on Clothing, where 46\% of users are text-weak, and by 1.3\% on MovieLens, where 27\% are (Tab.~\ref{tab:user_anchor_split}), so per-user identification is needed (case study in App.~\ref{app:case_study}).

\begin{table}[!t]
    \centering
    \caption{Penalty variants of $\mathcal{L}_{\text{WPCL}}$ in the VLM2Rec setting.}
    \label{tab:penalty_variants}
    \setlength{\tabcolsep}{9pt}
    \resizebox{\columnwidth}{!}{
    \renewcommand{\arraystretch}{0.8}
    \begin{tabular}{lcccc}
    \toprule
    \textbf{Penalty} & \textbf{Gap} & \begin{tabular}[c]{@{}c@{}}\textbf{Weak}\\\textbf{modality}\end{tabular} & \textbf{Clothing} & \textbf{MovieLens} \\
    \midrule
    None & -- & -- & 0.0342 & 0.1131 \\
    Static & -- & -- & 0.0347 & 0.1145 \\
    Batch-level & Batch & Batch & 0.0352 & 0.1212 \\
    Modality-fixed & User & Vision & 0.0363 & 0.1236 \\
    \textbf{User-adaptive} & User & User & \textbf{0.0370} & \textbf{0.1252} \\
    \bottomrule
    \end{tabular}
    }
\end{table}

\begin{figure}[t]
    \centering
    \includegraphics[width=\columnwidth]{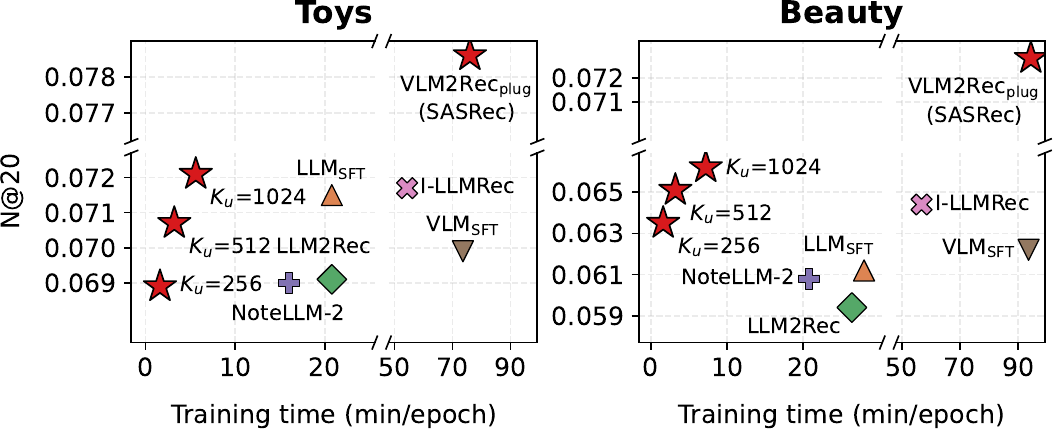}
    \Description{Two scatter panels, Toys and Beauty, plotting training time per epoch against N@20 with broken axes. VLM2Rec-plug with SASRec trained on all users is the highest point. Its variants trained on 256, 512, and 1024 users take a few minutes per epoch, and the 1024-user variant outperforms every baseline in both panels.}
    \caption{Few-shot efficiency: training time (min/epoch).}
    \label{fig:few_shot}
\end{figure}

\subsection{Generalization under Scarce CF Signals}

We evaluate generalization under scarce CF signals in two settings: \textit{cold-start items} (Fig.~\ref{fig:scarce_cf}a) and \textit{cross-domain transfer} (Fig.~\ref{fig:scarce_cf}b).
\noindent\textbf{Cold-start Analysis}: 
Items are grouped by their training interactions (0, 1--10, 11--30, and 31+), and new items are ranked from their titles and images alone.
Rarer items leave fewer transitions for the backbone to memorize, so ranking them relies more on content semantics; accordingly, $\text{VLM2Rec}_{\text{plug}}$'s relative gain over the ID-only SASRec is largest for items with at most ten training interactions.
\noindent\textbf{Cross-domain Transfer}: The embedder trained on a source domain encodes target-domain items for an SR backbone trained and evaluated on that target; domain shift makes memorized CF signals unreliable.
Here, the multimodal embedders, such as I-LLMRec, transfer better than the text-only $\text{LLM}_{\text{SFT}}$, suggesting that visual preferences carry across domains.
$\text{VLM2Rec}_{\text{plug}}$ achieves the best results in both settings, suggesting that balanced multimodal representations transfer well when CF signals are scarce.

\subsection{Cost and Few-shot Efficiency Study}
\label{sec:few_shot}

Fig.~\ref{fig:few_shot} reports training time (min/epoch) across LLM- and VLM-based methods, and Tab.~\ref{tab:emb_time} reports item embedding generation time.
$\text{VLM}_{\text{SFT}}$ and $\text{VLM2Rec}_{\text{plug}}$ are more expensive to train than other methods because they process image sequence tokens, but their similar runtimes show that this cost arises from VLM encoding rather than our losses.
To reduce this cost, we train a few-shot variant on $K_u$ random users: at $K_u{=}1024$ (about 5\% of training data), it already exceeds the strongest baseline with much shorter training time.
The complexity and hyperparameter analyses confirm that our losses preserve the asymptotic complexity of $\mathcal{L}_{\text{SFT}}$ (Tab.~\ref{tab:complexity_loss}) and that performance is stable across hyperparameters (Fig.~\ref{fig:hyperparameter}).
Performance also grows with VLM size and is similar across families (Fig.~\ref{fig:model_capacity}), so the VLM can be scaled to the compute budget without changing the serving cost of $\text{VLM2Rec}_{\text{plug}}$.

\section{Conclusion}
\label{sec:conclusion}
We presented VLM2Rec, a framework that adapts VLMs into CF-aware multimodal sequence embedders for SR.
Our analysis shows that SFT amplifies the pre-existing modality collapse of VLMs, suppressing the weak modality and turning visual information into negative transfer.
VLM2Rec mitigates this collapse, and its variant $\text{VLM2Rec}_{\text{plug}}$ yields consistent gains across datasets, SR backbones, and VLM families without adding serving latency.

\bibliographystyle{ACM-Reference-Format}
\bibliography{reference}

\appendix
\begin{table}[t]
\centering
\caption{Statistics of datasets used in the experiments.}
\renewcommand{\arraystretch}{0.8}
\resizebox{\linewidth}{!}{
\begin{tabular}{c|ccccc}
\toprule
\textbf{Dataset} & \textbf{\#Users} & \textbf{\#Items} & \textbf{\#Interactions} & \textbf{Avg. Length} & \textbf{Sparsity (\%)} \\
\midrule
Toys  & 19,406 & 11,842 & 166,230 & 8.6   & 99.93 \\
Beauty  & 22,363 & 12,077 & 198,064 & 8.9   & 99.93 \\
Sports  & 35,597 & 18,168 & 294,045  & 8.3   & 99.95 \\
Clothing   & 39,387 & 22,445 & 270,319  & 6.9   & 99.97 \\
MovieLens  & 6,040 & 3,397 & 996,867  & 165   & 95.14 \\
\bottomrule
\end{tabular}
}
\label{tab:dataset_stats}
\end{table}
\begin{figure}[!t]
    \centering
    \includegraphics[width=\columnwidth]{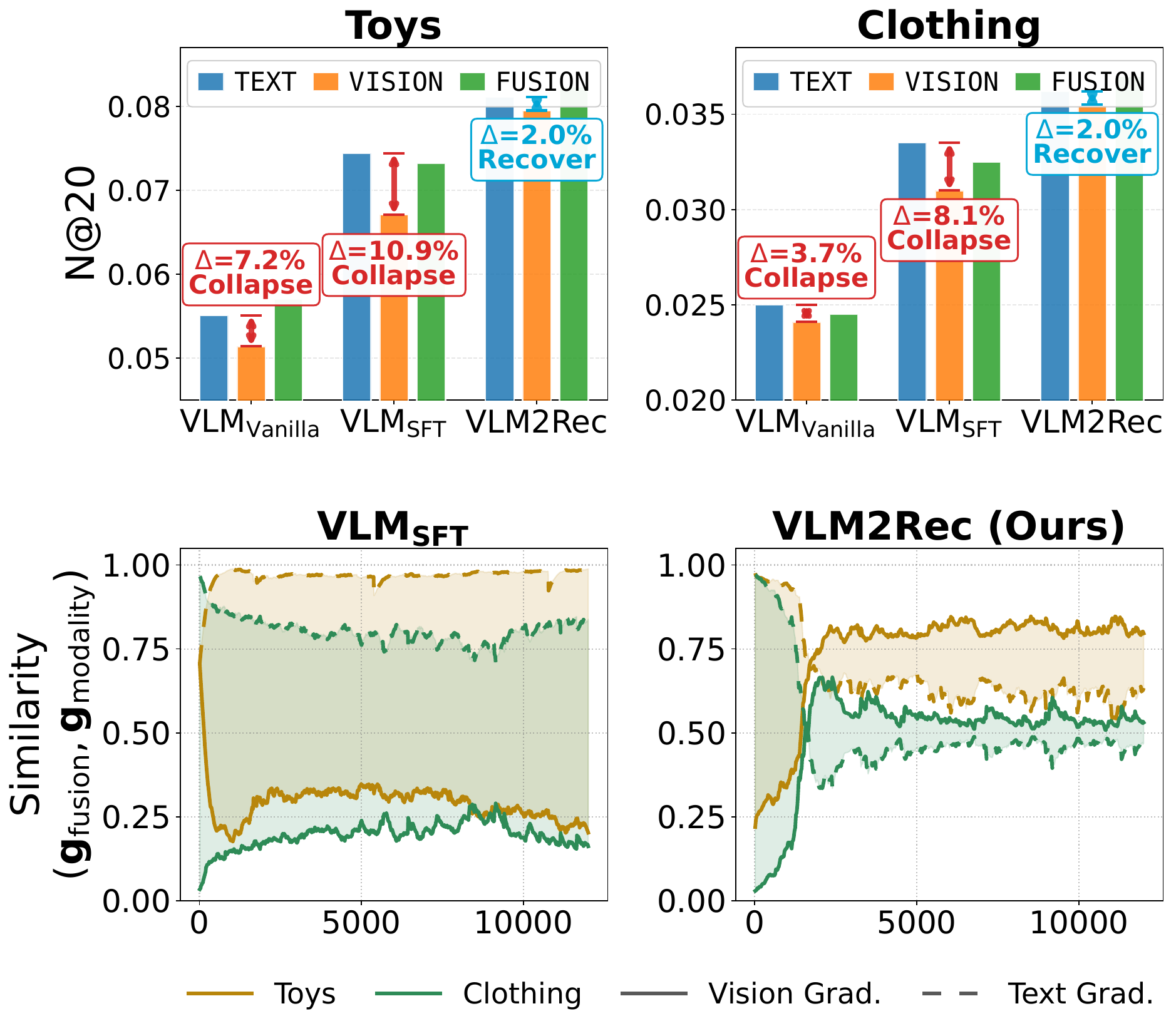}
    \Description{Bar charts compare modality-specific and fused recommendation performance on Toys and Clothing. Line charts show text and vision gradient alignment during SFT and VLM2Rec training.}
    \caption{Modality collapse on Toys and Clothing: dropout test (top) and gradient dynamics (bottom).}
    \label{fig:drop_grad_toys_clothing}
\end{figure}


\section{Implementation Details}
\label{app:exp_details}

\noindent\textbf{Datasets.}
\label{app:datasets}
We remove items lacking a title or an image (Tab.~\ref{tab:dataset_stats}); for MovieLens, we collect posters from TMDB using the IMDb and TMDB IDs in the MovieLens links file.

\vspace{+0.1cm}
\noindent\textbf{VLM2Rec Pretraining.}
\label{app:implementation_details}
We apply LoRA~\cite{lora} (rank 16, alpha 32, dropout 0.2) and train for 3 epochs with AdamW (learning rate $1\text{e-}5$, batch size 32).
Grid search selects
$\tau_{\text{WPCL}} \in \{0.05, 0.1, 0.5, 1.0\}$,
$\tau_{\text{CRTR}} \in \{0.001, 0.01, 0.1, 1.0\}$, and
$\lambda \in \{0.1, 0.5, 1.0\}$.
With gradient checkpointing, experiments run on one RTX 3090 by default and on one A100 80GB GPU for larger models.

\vspace{+0.1cm}
\noindent\textbf{SR Backbone Training.}
SR backbones are trained on a single RTX 3090 with Adam (learning rate 0.001, $\beta_1=0.9$, $\beta_2=0.999$), batch size 256, dropout 0.5, and early stopping with a patience of 10 epochs on validation N@20.
Baseline hyperparameters are tuned over the search spaces of their original papers.

\vspace{+0.1cm}
\noindent\textbf{Text-based Baselines.}
We use BERT\footnote{google-bert/bert-large-uncased} for text-only item encoding; it embeds items independently, carries no sequence-level CF, and forms sequence representations by mean-pooling its item embeddings.
Following its original paper, SLIM extracts final representations with BERT rather than its student LLM distilled from the ChatGPT API\footnote{gpt-4.1-mini}, so its CF-trained foundation model does not produce the embeddings; our $\text{SLIM}^{+}$ instead extracts them from the student, and both are trained for next-item generation rather than for embedding quality.
LLM2Rec injects sequence-level CF into a text-only LLM embedder, whereas LLMEmb injects CF only indirectly by distilling ID embeddings.
$\text{LLM}_{\text{Vanilla}}$ directly extracts sequence and item embeddings from the foundation model without further training, while $\text{LLM}_{\text{SFT}}$ applies sequence-level contrastive fine-tuning (Eq.~\ref{eq:sft_loss}).

\vspace{+0.1cm}
\noindent\textbf{Multimodal-based Baselines.}
We use CLIP\footnote{openai/clip-vit-large-patch14} for multimodal item encoding, pooled in the same way as BERT.
NoteLLM-2 learns multimodal item embeddings without sequence-level CF, fusing vision through a simple gated sum rather than a dedicated modality-balancing design, and $\text{VLM}_{\text{Prompt}}$ obtains item embeddings by zero-shot prompting without tuning.
I-LLMRec feeds each item's CLIP image feature into a text-only LLM as a single input token, following its original design.
For comparison under the same protocol, we replace the yes/no prediction loss of MLLM-MSR with standard SFT (Eq.~\ref{eq:sft_loss}).
$\text{VLM}_{\text{Vanilla}}$ and $\text{VLM}_{\text{SFT}}$ are the VLM counterparts of $\text{LLM}_{\text{Vanilla}}$ and $\text{LLM}_{\text{SFT}}$; $\text{VLM}_{\text{SFT}}$ thus shares VLM2Rec's backbone, inputs, and CF injection, lacking only modality balancing.


\begin{table}[!t]
\centering
\caption{VLM geometry on Toys and Clothing: uniformity $U$ and separability $S$.}
\label{tab:ali_uni_toys_clothing}
\setlength{\tabcolsep}{2pt}
\renewcommand{\arraystretch}{0.5}
\resizebox{\columnwidth}{!}{%
\begin{tabular}{@{}lccccccc@{}}
\toprule
\multirow{2}{*}{\textbf{Dataset}} & \multirow{2}{*}{\textbf{Mod.}} & \multicolumn{3}{c}{$U$ ($\downarrow$)} & \multicolumn{3}{c}{$S$ ($\uparrow$)} \\
\cmidrule(lr){3-5}\cmidrule(l){6-8}
 & & $\text{VLM}_{\text{Vanilla}}$ & $\text{VLM}_{\text{SFT}}$ & \textbf{Ours} & $\text{VLM}_{\text{Vanilla}}$ & $\text{VLM}_{\text{SFT}}$ & \textbf{Ours} \\
\midrule
\multirow{3}{*}{\textbf{Toys}}
 & Fused & -0.4707 & -2.2500 & -1.7969 & 1.0609 & 1.2418 & 1.1609 \\
 & Vision & \textbf{-0.0237} & \textcolor{red}{\textbf{-0.0039}} & \textcolor{blue}{\textbf{-0.5703}} & \textbf{1.0039} & \textcolor{red}{\textbf{0.9911}} & \textcolor{blue}{\textbf{1.1551}} \\
 & Text & -0.4453 & -2.2188 & -0.9335 & 1.0673 & 1.2480 & 1.1218 \\
\midrule
\multirow{3}{*}{\textbf{Clothing}}
 & Fused & -0.4336 & -1.8359 & -1.9951 & 1.0444 & 1.0474 & 1.1274 \\
 & Vision & \textbf{-0.2368} & \textcolor{red}{\textbf{-0.0237}} & \textcolor{blue}{\textbf{-1.2136}} & \textbf{1.0110} & \textcolor{red}{\textbf{0.9997}} & \textcolor{blue}{\textbf{1.0706}} \\
 & Text & -0.4102 & -1.7187 & -1.3759 & 1.0551 & 1.0562 & 1.0610 \\
\bottomrule
\end{tabular}%
}
\end{table}

\begin{table}[!t]
\centering
\caption{Frozen-VLM Head-SFT control in the $\text{VLM2Rec}_{\text{plug}}$ setting (SASRec, N@20).}
\label{tab:head_sft_control}
\setlength{\tabcolsep}{3pt}
\renewcommand{\arraystretch}{0.8}
\begin{tabular*}{\columnwidth}{@{\extracolsep{\fill}}llcccc@{}}
\toprule
\textbf{Dataset} & \textbf{Variant} & \textbf{\TEXTset} & \textbf{\VISIONset} & \textbf{\FUSIONset} & \textbf{Gap} (\%) \\
\midrule
\multirow{3}{*}{Beauty}
 & $\text{VLM}_{\text{Vanilla}}$ & 0.0469 & 0.0438 & 0.0463 & 7.08 \\
 & Head-SFT    & 0.0467 & 0.0385 & 0.0462 & 21.30 \\
 & $\text{VLM}_{\text{SFT}}$     & 0.0650 & 0.0587 & 0.0622 & 10.73 \\
\midrule
\multirow{3}{*}{MovieLens}
 & $\text{VLM}_{\text{Vanilla}}$ & 0.0665 & 0.0622 & 0.0695 & 6.91 \\
 & Head-SFT    & 0.0950 & 0.0751 & 0.0916 & 26.50 \\
 & $\text{VLM}_{\text{SFT}}$     & 0.1100 & 0.0953 & 0.1082 & 15.42 \\
\bottomrule
\end{tabular*}
\end{table}

\section{Additional Analysis}
\label{app:additional_empirical_analyses}

\subsection{Modality-collapse Analyses}

\noindent\textbf{Additional Domains.}
\label{app:additional_analysis}
Standard SFT again biases the fused update toward the stronger modality and widens the \TEXTset{}--\VISIONset{} gap on Toys and on Clothing (Fig.~\ref{fig:drop_grad_toys_clothing}), so the pattern holds even in the visually informative Clothing domain.
The geometry repeats the pattern of Tab.~\ref{tab:ali_uni_clo} (Tab.~\ref{tab:ali_uni_toys_clothing}): $\text{VLM}_{\text{SFT}}$ keeps visual separability $S$ at or below the discrimination threshold and degrades visual uniformity $U$, whereas VLM2Rec restores both, balances the gradients, and lets \FUSIONset{} surpass either single-modality input, consistent with Sec.~\ref{sec:modality_mute}.

\begin{table}[!t]
\centering
\caption{Performance on additional SR backbones (N@20). \protect\colorbox{bestcolor}{\textbf{Bold}}/\protect\colorbox{secondcolor}{\underline{underline}}: best/second best; *: $p<0.05$ (paired $t$-test, 5 seeds).}
\setlength{\tabcolsep}{3pt}
\resizebox{\columnwidth}{!}{
\renewcommand{\arraystretch}{0.8}
\begin{tabular}{c|cc|cccc|cc}
\toprule
\multirow{2}{*}{\textbf{Data}} & \multicolumn{2}{c|}{\textbf{Text-based Models}} & \multicolumn{4}{c|}{\textbf{Multimodal-based Models}} & \multicolumn{2}{c}{\textbf{Ours}} \\
\cmidrule(lr){2-3} \cmidrule(lr){4-7} \cmidrule(lr){8-9}
 & LLM2Rec & $\text{LLM}_{\text{SFT}}$ & $\text{VLM}_{\text{Van}}$ & $\text{VLM}_{\text{SFT}}$ & MLLM-MSR & I-LLMRec & \textbf{$\text{VLM2Rec}_{\text{plug}}$} & \textit{Imp(\%)} \\ \midrule

\midrule[\heavyrulewidth]
\multicolumn{9}{c}{\cellcolor{gray!15}\textbf{Backbone: DiffuRec}} \\ \midrule

\textbf{Toys} & 0.0639 & 0.0644 & 0.0586 & 0.0655 & 0.0650 & \secondscore{0.0658} & \bestscore{0.0760*} & 15.50 \\

\textbf{Beauty} & 0.0656 & 0.0670 & 0.0561 & 0.0682 & 0.0674 & \secondscore{0.0689} & \bestscore{0.0740*} & 7.40 \\

\textbf{Clothing} & 0.0316 & 0.0325 & 0.0291 & \secondscore{0.0334} & 0.0323 & 0.0327 & \bestscore{0.0384*} & 14.97 \\

\textbf{Sports} & 0.0319 & 0.0334 & 0.0288 & 0.0322 & 0.0340 & \secondscore{0.0347} & \bestscore{0.0398*} & 14.70 \\

\textbf{MovieLens} & 0.1036 & 0.1071 & 0.0767 & 0.1096 & 0.1077 & \secondscore{0.1112} & \bestscore{0.1188*} & 6.83 \\
\midrule[\heavyrulewidth]
\multicolumn{9}{c}{\cellcolor{gray!15}\textbf{Backbone: FMLP}} \\ \midrule

\textbf{Toys} & 0.0639 & 0.0660 & 0.0534 & 0.0642 & 0.0664 & \secondscore{0.0670} & \bestscore{0.0732*} & 9.25 \\

\textbf{Beauty} & 0.0552 & 0.0564 & 0.0432 & 0.0585 & 0.0569 & \secondscore{0.0598} & \bestscore{0.0691*} & 15.55 \\

\textbf{Clothing} & 0.0265 & 0.0269 & 0.0206 & 0.0281 & 0.0264 & \secondscore{0.0285} & \bestscore{0.0341*} & 19.65 \\

\textbf{Sports} & 0.0272 & 0.0278 & 0.0225 & 0.0284 & 0.0280 & \secondscore{0.0285} & \bestscore{0.0353*} & 23.86 \\

\textbf{MovieLens} & 0.0844 & 0.0867 & 0.0582 & \secondscore{0.0913} & 0.0876 & 0.0909 & \bestscore{0.1057*} & 15.77 \\
\midrule[\heavyrulewidth]
\multicolumn{9}{c}{\cellcolor{gray!15}\textbf{Backbone: GRU4Rec}} \\ \midrule

\textbf{Toys} & 0.0447 & 0.0454 & 0.0424 & 0.0465 & 0.0463 & \secondscore{0.0475} & \bestscore{0.0576*} & 21.26 \\

\textbf{Beauty} & 0.0470 & 0.0476 & 0.0422 & 0.0485 & 0.0478 & \secondscore{0.0491} & \bestscore{0.0548*} & 11.61 \\

\textbf{Clothing} & 0.0231 & 0.0233 & 0.0204 & \secondscore{0.0242} & 0.0231 & 0.0239 & \bestscore{0.0264*} & 9.09 \\

\textbf{Sports} & 0.0267 & 0.0279 & 0.0216 & 0.0273 & 0.0283 & \secondscore{0.0289} & \bestscore{0.0305*} & 5.54 \\

\textbf{MovieLens} & 0.0829 & 0.0867 & 0.0587 & \secondscore{0.0902} & 0.0876 & 0.0896 & \bestscore{0.0965*} & 6.98 \\\bottomrule

 \end{tabular}
}
\label{tab:add_backbones}
\end{table}

\noindent\textbf{Analysis of the SFT Objective.}
\label{app:head_sft_control}
Fused contrastive SFT can amplify modality collapse with a frozen VLM (Tab.~\ref{tab:head_sft_control}).
To separate its effect from LoRA updates (Fig.~\ref{fig:drop_grad}), we train only a linear head with the same loss (Head-SFT), then evaluate the item embeddings with SASRec in the $\text{VLM2Rec}_{\text{plug}}$ setting.
The \TEXTset{}--\VISIONset{} gap widens beyond that of $\text{VLM}_{\text{Vanilla}}$ on both datasets, and head-only gradients remain text-dominant.

\subsection{Robustness and Efficiency}

\noindent\textbf{Additional Backbone Results.}
\label{sec:add_backbone}
$\text{VLM2Rec}_{\text{plug}}$ performs best on the diffusion-based DiffuRec~\cite{diffurec}, the MLP-based FMLP~\cite{fmlp}, and the RNN-based GRU4Rec~\cite{gru4rec} (Tab.~\ref{tab:add_backbones}), indicating that its embeddings transfer across architectures.

\begin{table}[!htb]
\centering
\caption{Modality dropout test across VLM backbones (Beauty, SASRec, N@20).}
\label{tab:modality_collapse_backbones}
\setlength{\tabcolsep}{2.5pt}
\resizebox{\columnwidth}{!}{
\renewcommand{\arraystretch}{0.8}
\begin{tabular}{l|ccc|c|ccc|c}
\toprule
\multirow{2}{*}{\textbf{VLM Backbone}}
 & \multicolumn{4}{c|}{\textbf{$\text{VLM}_{\text{SFT}}$}}
 & \multicolumn{4}{c}{\textbf{VLM2Rec}} \\
\cmidrule(lr){2-5} \cmidrule(lr){6-9}
 & \TEXTset{} & \VISIONset{} & \FUSIONset{} & \textbf{Gap (\%)}
 & \TEXTset{} & \VISIONset{} & \FUSIONset{} & \textbf{Gap (\%)} \\
\midrule
InternVL3-2B    & 0.0614 & 0.0558 & 0.0608 & 10.04 & 0.0703 & 0.0686 & 0.0702 & \textbf{2.48} \\
Qwen2.5-VL-7B   & 0.0636 & 0.0588 & 0.0632 &  8.16 & 0.0746 & 0.0733 & 0.0754 & \textbf{1.77} \\
LLaVA-1.5-7B    & 0.0625 & 0.0572 & 0.0620 &  9.27 & 0.0733 & 0.0725 & 0.0742 & \textbf{1.10} \\
InternVL3-8B    & 0.0645 & 0.0597 & 0.0636 &  8.04 & 0.0744 & 0.0727 & 0.0750 & \textbf{2.34} \\
Qwen2.5-VL-32B  & 0.0653 & 0.0612 & 0.0647 &  6.70 & 0.0761 & 0.0744 & 0.0776 & \textbf{2.28} \\
\bottomrule
\end{tabular}
}
\end{table}

\begin{figure}[!htb]
    \centering
    \includegraphics[width=\columnwidth]{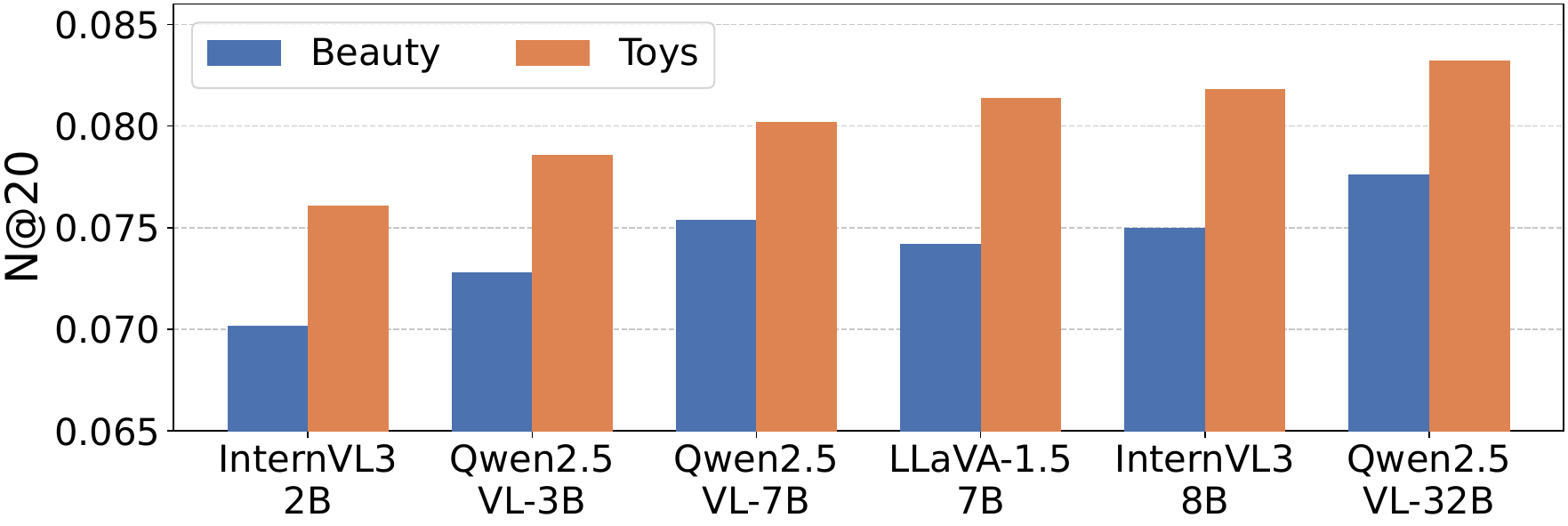}
    \Description{Grouped bars compare VLM2Rec's N@20 on Beauty and Toys for six VLM backbones from three families, ranging from two billion to thirty-two billion parameters.}
    \caption{Model scalability across VLM families and parameter sizes (SASRec, N@20).}
    \label{fig:model_capacity}
\end{figure}

\noindent\textbf{VLM Backbones and Scalability.}
\label{app:modality_collapse_backbones}
\label{app:model_scalability}
We rerun the modality dropout test on Beauty across five backbones spanning three model families (InternVL3~\cite{internvl3}, Qwen2.5-VL~\cite{qwen2.5-vl}, LLaVA-1.5~\cite{llava}) and four scales (2B--32B) (Tab.~\ref{tab:modality_collapse_backbones}).
For every backbone, $\text{VLM}_{\text{SFT}}$ shows a sizable \TEXTset{}--\VISIONset{} gap, which VLM2Rec substantially narrows while improving \FUSIONset{}, so the SFT-amplified collapse and its mitigation are not artifacts of a specific backbone.
VLM2Rec's recommendation performance also scales with model capacity on both Beauty and Toys, while models of similar size perform comparably across families (Fig.~\ref{fig:model_capacity}), so the model size can follow deployment budgets.

\begin{table}[!h]
\centering
\caption{VLM2Rec on Qwen3-VL-Embedding in the $\text{VLM2Rec}_{\text{plug}}$ setting (SASRec, N@20). Base: the released model without our training. Gap: (\TEXTset{}$-$\VISIONset{})/\VISIONset{}; lower is better.}
\label{tab:qwen3_vl_embedding}
\setlength{\tabcolsep}{7pt}
\renewcommand{\arraystretch}{0.5}
\resizebox{\linewidth}{!}{%
\begin{tabular}{lllcccc}
\toprule
\textbf{Data} & \textbf{Size} & \textbf{Variant} & \textbf{\TEXTset} & \textbf{\VISIONset} & \textbf{\FUSIONset} & \textbf{Gap (\%)} \\
\midrule
Beauty & 2B & Base & 0.0732 & 0.0672 & 0.0719 & 8.93 \\
       &    & +VLM2Rec & 0.0751 & 0.0718 & 0.0746 & 4.60 \\
       & 8B & Base & 0.0745 & 0.0692 & 0.0729 & 7.66 \\
       &    & +VLM2Rec & 0.0762 & 0.0734 & 0.0751 & 3.81 \\
\midrule
MovieLens & 2B & Base & 0.1176 & 0.1034 & 0.1147 & 13.73 \\
          &    & +VLM2Rec & 0.1179 & 0.1098 & 0.1162 & 7.38 \\
          & 8B & Base & 0.1211 & 0.1119 & 0.1186 & 8.22 \\
          &    & +VLM2Rec & 0.1206 & 0.1187 & 0.1192 & 1.60 \\
\bottomrule
\end{tabular}%
}
\end{table}

\noindent\textbf{Pretrained Multimodal Embedders.}
\label{app:embedding_vlm_transfer}
We apply VLM2Rec to the open-source Qwen3-VL-Embedding 2B and 8B models~\cite{qwen3_vl_embedding}, already trained on large-scale multimodal retrieval tasks, on Beauty and MovieLens (Tab.~\ref{tab:qwen3_vl_embedding}).
Even these embedders show a sizable \TEXTset{}--\VISIONset{} gap (7.7--13.7\%).
Within each backbone, VLM2Rec substantially reduces the \TEXTset{}--\VISIONset{} gap and increases fused N@20 in all four comparisons, with the largest gains in vision-only ranking; text-only performance is not uniformly higher (e.g., the 8B MovieLens case).
These within-backbone results show that weak-modality correction also benefits pretrained multimodal embedders; we compare only within each backbone, since these embedders are trained at a far larger scale than the models in Tab.~\ref{tab:perf_task2}.

\begin{table}[!h]
\centering
\caption{Time and space complexity of the loss objectives.}
\renewcommand{\arraystretch}{0.8}
\setlength{\tabcolsep}{3pt}
\resizebox{\columnwidth}{!}{%
\begin{tabular}{lcccc}
\toprule
\textbf{Objective} & \textbf{Time} & \textbf{Space} & \textbf{In-batch Time} & \textbf{In-batch Space} \\
\midrule
$\mathcal{L}_{\text{SFT}}$
& $O\!\left(2B(N{+}1)d\right)$
& $O\!\left(B(N{+}1)\right)$
& $O(2B^2d)$
& $O(B^2)$ \\
$\mathcal{L}_{\text{WPCL}}$
& $O\!\left(6B(N{+}1)d\right)$
& $O\!\left(B(N{+}1)\right)$
& $O(6B^2d)$
& $O(B^2)$ \\
$\mathcal{L}_{\text{CRTR}}$
& $O(4BCd)$
& $O(BC)$
& $O(4B^2d)$
& $O(B^2)$ \\
\bottomrule
\end{tabular}%
}
\label{tab:complexity_loss}
\end{table}

\noindent\textbf{Computational Complexity.}
\label{app:complexity}
Tab.~\ref{tab:complexity_loss} compares the loss computation after backbone encoding, where $B = |\mathcal{B}|$ is the batch size, $N=|\mathcal{N}_u|$ the number of negatives per user, $C=|\mathcal{C}|$ the candidate set size of $\mathcal{L}_{\text{CRTR}}$, and $d$ the embedding dimension; space denotes the additional memory of the loss computation, excluding backbone activations.
$\mathcal{L}_{\text{WPCL}}$ adds two modality-wise similarity passes to $\mathcal{L}_{\text{SFT}}$ with the same space, and $\mathcal{L}_{\text{CRTR}}$ builds modality-wise sequence-to-candidate similarity matrices.
With in-batch negatives and candidates ($N=B-1$, $|\mathcal{C}|=B$), all three losses cost $O(B^2d)$ time and $O(B^2)$ space, differing only in constant factors.

\begin{table}[t]
\centering
\caption{Item embedding generation time (sec), measured once per catalog.}
\label{tab:emb_time}
\setlength{\tabcolsep}{3pt}
\resizebox{\columnwidth}{!}{
\renewcommand{\arraystretch}{0.7}
\begin{tabular}{l|cccccc|c}
\toprule
\textbf{Dataset} & LLMEmb & LLM2Rec & $\text{LLM}_{\text{SFT}}$ & NoteLLM-2 & $\text{VLM}_{\text{SFT}}$ & I-LLMRec & $\text{VLM2Rec}_{\text{plug}}$ \\
\midrule
Beauty & 58 & 58 & 56 & 132 & 113 & 60 & 113 \\
Toys & 38 & 39 & 39 & 91 & 79 & 41 & 79 \\
\bottomrule
\end{tabular}
}
\end{table}

\vspace{+0.1cm}
\noindent\textbf{Item Embedding Generation Time.}
\label{app:emb_time}
Tab.~\ref{tab:emb_time} reports the time to encode each catalog once, complementing the training times in Fig.~\ref{fig:few_shot}.
$\text{VLM2Rec}_{\text{plug}}$ takes the same time as $\text{VLM}_{\text{SFT}}$ for every $K_u$, since both encode titles and images with the same VLM.

\begin{table}[t]

\centering
\caption{Effect of negative sampling strategies for $\Delta_{u,\mathrm{gap}}$ estimation (SASRec, N@20).}
\label{tab:negative_sampling}
\resizebox{\linewidth}{!}{
\renewcommand{\arraystretch}{0.7}
\begin{tabular}{l|ccccc}
\toprule
\textbf{Dataset} & \textbf{In-batch} & \textbf{Random} & \textbf{SASRec Top-$k$} & \textbf{Text Sim.} & \textbf{Image Sim.} \\
\midrule
Clothing & 0.0345 & 0.0342 & 0.0352 & 0.0357 & 0.0360 \\
MovieLens    & 0.1216 & 0.1193 & 0.1218 & 0.1222 & 0.1229 \\
\bottomrule
\end{tabular}
}

\end{table}

\vspace{+0.1cm}
\noindent\textbf{Negative Sampling for $\Delta_{u,\mathrm{gap}}$ Estimation.}
\label{app:negative_sampling}
Harder negatives for estimating $\Delta_{u,\text{gap}}$ (Eq.~\ref{eq:modality_gap}) improve N@20 over the in-batch negatives of our main experiments (Tab.~\ref{tab:negative_sampling}).
SASRec Top-$k$ uses each user's top-$k$ SASRec predictions, while Text Sim. and Image Sim. sample target-similar items by BERT and CLIP similarity, respectively.
Image-similar negatives perform best, suggesting that fine-grained visual distinctions sharpen the $\Delta_{u,\mathrm{gap}}$ estimate for the weak visual modality; we keep in-batch negatives to avoid retrieval cost, while modality-aware hard negatives could help further.

\begin{figure}[!h]
    \centering
    \includegraphics[width=\linewidth]{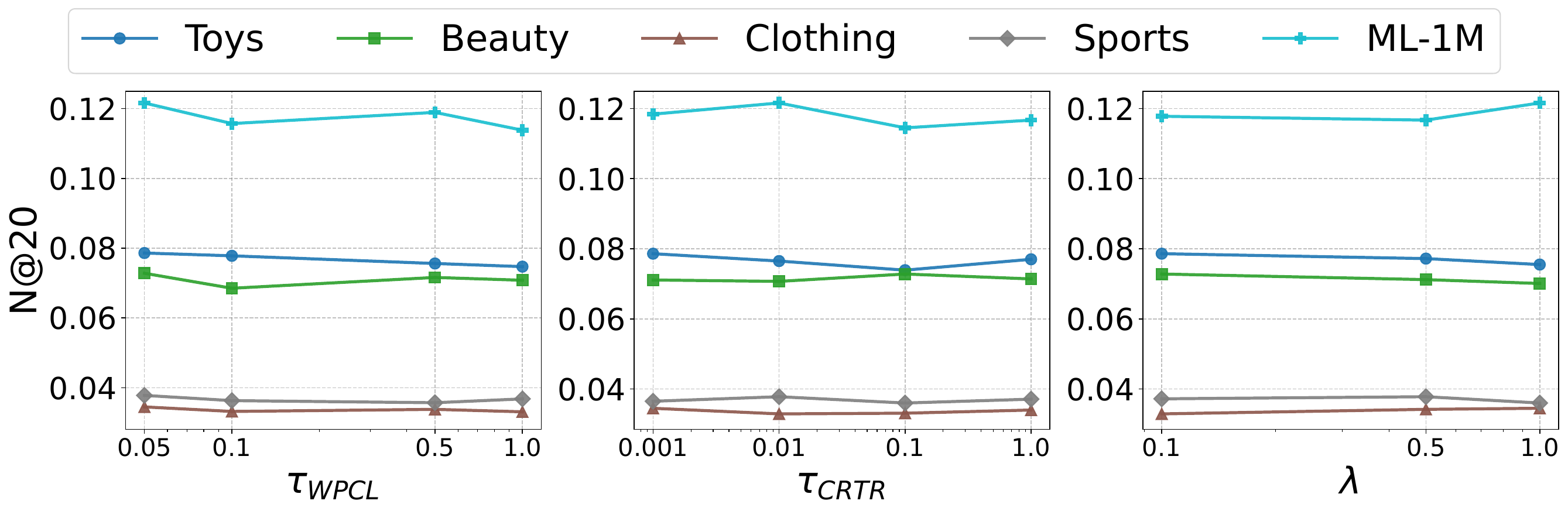}
    \Description{Three line plots show VLM2Rec-plug N@20 with SASRec on five datasets as the WPCL temperature, CRTR temperature, and regularization weight vary.}
    \caption{Hyperparameter study of $\text{VLM2Rec}_{\text{plug}}$ (SASRec, N@20).}
    \label{fig:hyperparameter}
\end{figure}

\noindent\textbf{Hyperparameter Study.}
\label{app:hyperparameter}
Across the tested settings, N@20 varies little and stays above the baselines (Fig.~\ref{fig:hyperparameter}).
This limits the tuning cost of $\text{VLM2Rec}_{\text{plug}}$.

\subsection{User-level Analysis}

\begin{table}[!t]
\centering
\caption{Anchor weak-modality split. Ratio: fraction of users whose anchor weak modality is text or vision.}
\label{tab:user_anchor_split}
\resizebox{\linewidth}{!}{
\renewcommand{\arraystretch}{0.7}
\begin{tabular}{lllcccc}
\toprule
\textbf{Data} & \textbf{Group} & \textbf{Weak} & \textbf{Ratio} & \textbf{$\Delta_{u,\mathrm{gap}}^{\text{SFT}}$} & \textbf{$\Delta_{u,\mathrm{gap}}^{\text{Ours}}$} & \textbf{User Red.} \\
\midrule
\multirow{4}{*}{MovieLens}
 & \multirow{2}{*}{All}      & Text   & 0.2728 & 0.2864 & 0.0235 & 0.9064 \\
 &                            & Vision & 0.7272 & 0.4047 & 0.0639 & 0.9473 \\
\cmidrule(lr){2-7}
 & \multirow{2}{*}{Collapse} & Text   & 0.2090 & 0.5563 & 0.0250 & 1.0000 \\
 &                            & Vision & 0.7910 & 0.6522 & 0.0808 & 1.0000 \\
\midrule
\multirow{4}{*}{Clothing}
 & \multirow{2}{*}{All}      & Text   & 0.4621 & 0.1346 & 0.0201 & 0.8224 \\
 &                            & Vision & 0.5379 & 0.1467 & 0.0232 & 0.8326 \\
\cmidrule(lr){2-7}
 & \multirow{2}{*}{Collapse} & Text   & 0.4891 & 0.2797 & 0.0529 & 0.9923 \\
 &                            & Vision & 0.5109 & 0.2803 & 0.0592 & 0.9932 \\
\bottomrule
\end{tabular}
}
\end{table}

\noindent\textbf{Anchor Weak-modality Analysis.}
\label{app:anchor_weak_split}
We split each user group by the anchor weak modality, the weaker modality under $\text{VLM}_{\text{Vanilla}}$; \textit{All} and \textit{Collapse} follow Sec.~\ref{sec:user_stratification}.
Vision is the weak modality for most users on both datasets (Tab.~\ref{tab:user_anchor_split}), consistent with the text dominance of VLMs, although its share drops on the visually rich Clothing.
VLM2Rec substantially lowers the gap from $\Delta_{u,\mathrm{gap}}^{\text{SFT}}$ to $\Delta_{u,\mathrm{gap}}^{\text{Ours}}$ in every split, and most users show $\Delta_{u,\mathrm{gap}}^{\text{Ours}}<\Delta_{u,\mathrm{gap}}^{\text{SFT}}$ (\textit{User Red.}), confirming recovery for both text-weak and vision-weak users.

\begin{figure}[!t]
    \centering
    \includegraphics[width=\columnwidth]{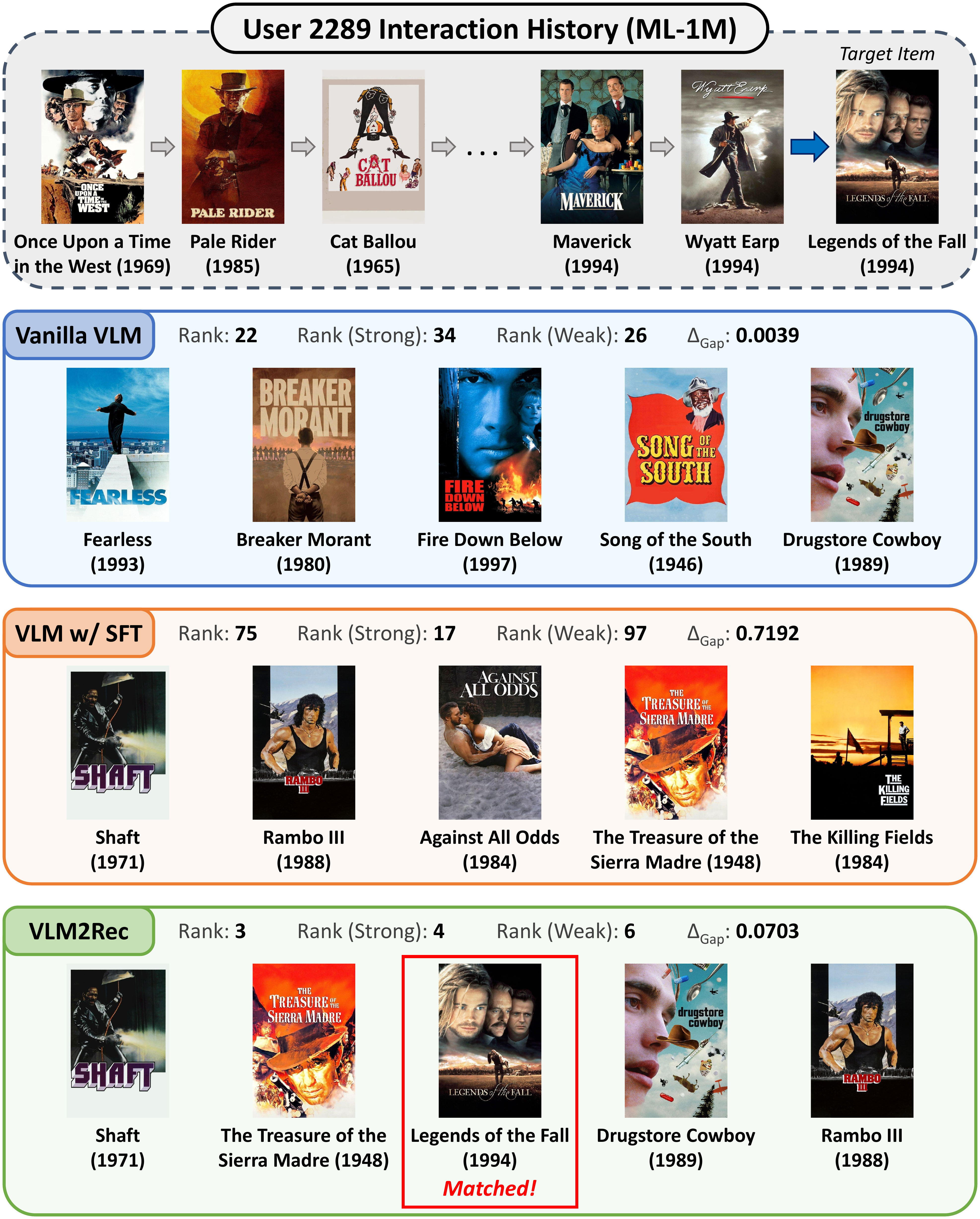}
    \Description{A MovieLens user history and target movie are followed by the top five recommendations from vanilla VLM, SFT, and VLM2Rec. The target rank worsens after SFT and improves with VLM2Rec.}
    \caption{Case study: Top-5 prediction lists of each model.}
    \label{fig:case_study}
\end{figure}

\vspace{+0.1cm}
\noindent\textbf{Case Study.}
\label{app:case_study}
For a \textit{Collapse} user (Sec.~\ref{sec:user_stratification}; Fig.~\ref{fig:case_study}), $\text{VLM}_{\text{SFT}}$'s recommendations are dominated by the stronger modality, illustrating how the weak-modality signal can be suppressed even after CF supervision is injected.
VLM2Rec instead recovers a more balanced multimodal representation and ranks candidates that better reflect the target preference, showing how weak-modality recovery translates into an individual recommendation.

\clearpage

\end{document}